\documentclass[10pt,a4paper]{article}
\pdfoutput=1
\usepackage{jheppub}
\usepackage{arydshln}
\usepackage{tikz}
\usetikzlibrary{positioning}
\usepackage{lipsum}
\usepackage{parskip}
\usepackage[all]{xy}
\usepackage{amsfonts}
\usepackage{amscd}
\usepackage{amssymb}
\usepackage{amsmath,bbm}
\usepackage{graphicx}
\usepackage{epsfig}
\usepackage{latexsym}
\usepackage{mathtools}
\usepackage{hyperref}
\usepackage{amsthm}
\usepackage[vcentermath]{youngtab}
\usepackage{accents}
    
\usepackage{calligra}

\DeclareMathAlphabet{\mathcalligra}{T1}{calligra}{m}{n}
\DeclareFontShape{T1}{calligra}{m}{n}{<->s*[2.2]callig15}{}

\def \be  {\begin{equation}}
\def \ee  {\end{equation}}
\def \bea {\begin{equation}\begin{aligned}}
\def \eea {\end{aligned}\end{equation}}
\def \ba  {\begin{eqnarray}}
\def \ea  {\end{eqnarray}}
\def \bb  {\begin{thebibliography}}
\def \eb  {\end{thebibliography}}
\def \lab #1 {\label{#1}}

\newcommand\cH{\mathcal{H}}

\newcommand\cT{\mathcal{T}}

\newcommand\qf{\mathfrak{q}}

\newcommand\bC{\mathbb{C }}

\newcommand\bR{\mathbb{R}}

\newcommand\bZ{\mathbb{Z}}

\renewcommand{\t}{\widetilde }

\definecolor{cardinal}{rgb}{0.6,0,0}
\definecolor{darkgreen}{rgb}{0,0.5,0}
\definecolor{golden}{rgb}{0.92, 0.7, 0}
\definecolor{midnight}{rgb}{0, 0, 0.5}
\definecolor{darkblue}{rgb}{0.2, 0, 0.8}

\theoremstyle{definition}

\def\CB{{\cal B}}

\def\CN{{\cal N}}

\def\CT{{\cal T}}

\def\g{\gamma}

\def\t{\widetilde}

\newcommand\wt{\widetilde}
\newcommand\bcT{\overline{\mathcal{T}}}
\newcommand\Tr{{\mathrm{Tr}}\,}
\newif\ifniklas\niklastrue
\RequirePackage{verbatim}

\makeatletter
\newwrite\bibinl@out
{\immediate\closeout\bibinl@out\@esphack}
\makeatother

\newcommand\fq{\mathfrak{q}}
\title{3D TFTs from 4d $\CN=2$ BPS Particles}
\author[1]{Davide Gaiotto,}
\affiliation[1]{Perimeter Institute for Theoretical Physicss, 31 Caroline Street North, Waterloo, ON N2L
2Y5, Canada}
\emailAdd{dgaiotto@perimeterinstitute.ca}
\author[2]{Heeyeon Kim}
\affiliation[2]{Department of Physics, Korea Advanced Institute of Science and Technology,Daejeon 34141, Republic of Korea}
\emailAdd{heeyeon.kim@kaist.ac.kr}

\abstract{We propose a general strategy to build three-dimensional gauge theories with four supercharges which enjoy a supersymmetry enhancement in the IR. The resulting IR SCFTs admit topological twists with particularly nice properties, as well as boundary rational chiral algebras such that the associated Modular Tensor Categories are controlled by the topological twist. The theories arise from a twisted circle compactification of four-dimensional theories of Argyres-Douglas type. We develop a novel algorithm to compute or manipulate protected quantities associated to these theories, such as ellipsoid partition functions and superconformal indices and half-indices. }
\begin{document}
	
\maketitle

\section{Introduction}

Three-dimensional Supersymmetric Quantum Field Theories with eight super-charges (${\cal N}=4$) provide a rich theoretical playground. A particularly interesting application is the relationship between the topological twist of these theories and certain 2d chiral algebras \cite{Costello:2018fnz}. This generalizes the foundational relation between unitary rational chiral algebras and unitary Topological Field Theories: the chiral algebras appear as edge modes for the corresponding TFTs. 

Fully supersymmetric Lagrangian descriptions of 3d ${\cal N}=4$ SQFTs have a very rigid structure and are fully determined by a choice of gauge group and matter representation. The corresponding TFTs and chiral algebras are complicated and very far from being rational. A remarkable recent development was the discovery of exotic 3d ${\cal N}=4$ SQFTs which do not admit such Lagrangian description and bypass the limitations associated to it \cite{Gang:2018huc,Gang:2021hrd,Gang:2023rei,Ferrari:2023fez,Creutzig:2024ljv,Gang:2023ggt,Baek:2024tuo}. In particular, this allows one to make contact with rational non-unitary chiral algebras such as ${\cal M}(p,q)$ Virasoro minimal models and the associated Modular Tensor Categories. 

The simplest exotic ${\cal N}=4$ theory $\cT_{\mathrm{min}}$ is defined as the IR limit of a 3d ${\cal N}=2$ $U(1)_{-\frac32}$ Chern-Simons theory coupled to a chiral multiplet of charge $1$ \cite{Gang:2018huc}. The theory has a $U(1)_A$ global symmetry which becomes part of the ${\cal N}=4$ R-symmetry group in the IR. The theory is expected to be mirror to the parity-reversed theory $\bcT_{\mathrm{min}}$ (i.e. the analogous $U(1)_{\frac32}$ CS theory). Correspondingly, the A- and B- twist give the TQFT associated to the Lee-Yang non-unitary minimal model ${\cal M}(2,5)$ or its parity conjugate \cite{Gang:2021hrd,Ferrari:2023fez}. 

The IR SUSY enhancement can be tested in multiple ways by computing protected quantities such as the ellipsoid partition function and the super-conformal index. It is also possible to define simple half-BPS boundary conditions and count BPS boundary operators via an half-index computation. If the boundary conditions are compatible with the topological twist, a specialization of the half-index will coincide with the vacuum character of the boundary chiral algebra. 

The half-index can be computed in the UV and the specialization gives rise to ``fermionic formulae'' or ``Nahm sum'' expressions for the chiral algebra characters:
\begin{equation}
    \chi_0[{\cal M}(2,5)](q) = \sum_{n=0}^\infty \frac{q^{n^2+n}}{(q;q)_n}
\end{equation}
For example, this sum arises from the half-index of a supersymmetric Dirichlet boundary for the gauge field in $\bar {\cal T}_{\mathrm{min}}$ (and a specific boundary condition for the matter fields). Each term in the sum counts boundary monopole operators of magnetic charge $n$. It can also be derived from a supersymmetric Neumann boundary condition for the gauge fields in ${\cal T}_{\mathrm{min}}$ (and a specific boundary condition for the matter fields).

General Nahm sum expressions 
\begin{equation}
    \sum_{n_i} \frac{q^{\frac12\vec n \cdot A \cdot \vec n + B \cdot \vec n +C}}{\prod_i (q;q)_{n_i}}
\end{equation}
exist for a large variety of other chiral algebras. Such an expression can be immediately identified with the specialized half-indices for supersymmetric boundary conditions in specific Abelian Chern-Simons theories, with a chiral multiplet of charge $1$ for each gauge field and a Chern-Simons level matrix computed from $A$. 

This observation was the starting point for the construction of more complicated examples of exotic 3d ${\cal N}=4$ theories \cite{Gang:2021hrd,Ferrari:2023fez,Creutzig:2024ljv}. A crucial step is the identification of appropriate superpotential deformation (often involving disorder operators) of the Abelian Chern-Simons theories derived from the Nahm sum, which  preserve an $U(1)_A$ symmetry generator which becomes part of the enhanced IR R-symmetry group but break other ``spurious'' symmetries. 

Remarkably, the same chiral algebra characters may admit multiple Nahm sum presentations. The corresponding superpotential-deformed Abelian CS (dACSM) theories  will be different, but may be IR dual and give rise to the same 3d ${\cal N}=4$ SCFTs. 

The same rational chiral algebras which appear in these constructions, such as non-unitary minimal models, also appear in a different physical context: protected operator algebras in 4d ${\cal N}=2$ SCFTs in the sense of \cite{Beem:2013sza}. The chiral algebra characters are identified with the so-called Schur-index of the 4d SCFTs. The two appearances are expected to be related by a super-symmetric cigar compactification \cite{Dedushenko:2023cvd}. In particular, the 3d ${\cal N}=4$ is expected to arise as a twisted circle compactification of the 4d theory. 

The relevant 4d theories are of Argyres-Douglas type and do not admit a traditional 4d ${\cal N}=2$ Lagrangian description. Some admit a 4d ${\cal N}=1$ Lagrangian description, which could be compactified to get some 3d ${\cal N}=2$ theories with expected IR enhancement \cite{Maruyoshi:2016tqk,Maruyoshi:2016aim,Agarwal:2016pjo,Dedushenko:2018bpp}. These theories are typically non-Abelian and do not lead to the same type of 3d theories deduced from fermionic formulae. 

The Schur index admits ``IR formulae'' involving the low-energy Seiberg-Witten description of the 4d theory and the spectrum of BPS particles \cite{Cordova:2015nma}. This paper elaborates on an observation from \cite{Cordova:2015nma}: the IR formulae also take the form of Nahm sums! Accordingly, they could be associated to auxiliary 3d dACS theories as well. 

We will argue that these 3d theories provide a natural dual description of
the twisted circle-compactified Argyres-Douglas theories, using the notions of theories of class $R$ \cite{Dimofte:2011py} and of R-flows \cite{Cecotti:2010fi,Cecotti:2011iy,Dimofte:2013lba,Cecotti:2015lab}.

We will also find systematic formulae to compute other protected quantities for such theories, such as ellipsoid partition functions and super-conformal indices. In particular, we will use ``Schur quantization'' \cite{Gaiotto:2024osr} to provide an universal test of the expected IR enhancement: a specialization of the index equals $1$. 

Demonstrating that the resulting dACSM theories are IR dual to the ones built from more conventional Nahm sum formulae is challenging but we will conduct some extensive tests. As an extra payoff, we will derive 3d theories associated to a larger variety of chiral algebras, some being non-rational in a controlled manner. 

Besides the three-dimensional applications, we hope that this construction will lead to SW-like constructions for the chiral algebras and for other protected quantities such as the category of line defects in an HT twist of the 4d SCFTs \cite{Kapustin:2006hi}, perhaps via a relation to Cohomological Hall Algebras \cite{Gaiotto:2024fso}.

\subsection{The main construction} 
The starting point of the construction is a 4d SCFT, with a relevant deformation which leads to the existence of a Coulomb vacuum where a Seiberg-Witten description is available and the spectrum of BPS particles is finite. 

The BPS spectrum then consists of a collection of $2N$ half-hypermultiplets of dyonic charges $\gamma_i$, which we order by the phase of their central charges $Z_{\gamma_i}$. The sequence includes both particles and anti-particles, so that $\gamma_{i+N} = - \gamma_i$.\footnote{As $Z_{-\gamma_i} = - Z_{\gamma_i}$.} 

The charges $\gamma_i$ are elements of the lattice $\Gamma$ of gauge and flavour charges in the IR Seiberg-Witten description. The charge lattice is equipped with an integral skew-symmetric Dirac pairing $\langle \gamma_i, \gamma_j\rangle$ which annihilates the sub-lattice $\Gamma_f$ of flavour charges. 

The resulting 3d theory is built from $2N$ 3d chiral multiplets.
We couple each chiral multiplet to an individual background $U(1)$ gauge field $A_i$ with $-\frac12$ units of Chern-Simons coupling, producing $2 N$ copies of the ``tetrahedron theory'' from \cite{Dimofte:2011ju}. We add to that further mixed CS couplings of the form 
\begin{equation}
    \sum_{i<j} \langle \gamma_i, \gamma_j\rangle A_i \wedge d A_j
\end{equation}
and gauge a subgroup $G$ of $U(1)^{2N}$ defined by the condition that 
\begin{equation}
    \sum_i A_i \gamma_i \in \Gamma_f
\end{equation}
Finally, we add a collection of super-potential terms, possibly involving monopole operators, which are discussed in greater detail in the main text. \footnote{Each undeformed ACSM theory belongs to a very large duality orbit of such theories. The original reference \cite{Dimofte:2011ju} defines theories of ``class R'' as dACSM theories such that each monomial in the super-potential should be a product of elementary fields. This condition allows for the application of a larger set of dualities and is rather natural in the current setup as well.}

We claim that this 3d theory has an IR SUSY enhancement and is dual to a twisted circle compactification of the original 4d theory. In particular, 3d theories associated to different chambers in the 4d Coulomb branch should belong to the same duality orbit. 

The construction can be generalized by repeating $k$ times the sequence of BPS particles to build a longer sequence. A small variant of the construction gives $k<0$. This is expected to produce other twisted circle compactification of the 4d theory, all enjoying IR SUSY enhancement. 

Deformed ACSM theories tend to have very large duality orbits.\footnote{Especially when they belong to class R, see previous footnote.} In concrete examples, we will attempt to find a duality representative which matches previous proposals in the literature. When we fail to do so, we will at least attempt to match some protected quantities. 

The construction is designed in such a way that the super-conformal half index counting protected operators for a specific boundary condition (for $k=1$) will match Nahm sum formulae for the Schur index of the 4d theory \cite{Cordova:2015nma}, i.e. the character of the associated chiral algebra. 

We will propose some remarkable formulae for the ellipsoid and super-conformal index of the 3d theory which have a ``shape'' analogous to the Nahm sums for the half-index: a ``trace'' of a product of $2 N$ quantum dilogarithms acting on auxiliary Hilbert spaces. One of the spaces is well-known from the theory of quantization of character varieties and the other from Schur quantization. 

The formulae may be of independent interest for other applications involving circle compactifications of 4d theories. They also generalize easily to correlation functions of BPS line defects. This may have applications to the computation of the modular matrix of the associated chiral algebras. 

\subsection{Note}
As this paper was being finalized we became aware of \cite{ardehali2024bridging}, which has some overlap with the results presented here.

\section{Abelian CSM theories and their boundary conditions}
The 3d ${\cal N}=2$ SQFTs we consider in this paper are Abelian Chern-Simons theories coupled to 3d chiral multiplets and deformed by super-potential terms which may include disorder operators: monopole operators dressed by chiral multiplet scalars. We refer to them as dACSM theories. The presence of disorder operators complicates the analysis of the systems, but many questions remain manageable. 

A basic subtlety concerns the quantization of the fermions in a chiral multiplet coupled to a background gauge field. In natural conventions, a 3d chiral multiplet effectively carries $-\frac12$ units of Chern-Simons coupling for the background gauge field coupled to it \cite{Seiberg:2016gmd}. This is referred to as the ``tetrahedron theory'' in \cite{Dimofte:2011ju}. In particular, parity maps such a 3d chiral multiplet to the same 3d chiral multiplet dressed by an extra $1$ unit of Chern-Simons coupling. We define the level matrix $K$ of the gauge fields in the theories as the CS couplings added on top of the ones associated to the chiral multiplets. 

In these conventions, for example, ${\cal T}_{\mathrm{min}}$ will be assigned $K=-1$ and ${\bcT}_{\mathrm{min}}$ will be assigned $K=2$. 
We will denote an $U(1)^N$ ACS theory with K-matrix K and $M$ chiral multiplets with a matrix of gauge charges $q$ as $\cT_{N,M}[K,q]$. The parity conjugate theory has a shifted $K$-matrix:
\begin{equation}
    \bcT_{N,M}[K,q] = \cT_{N,M}[q q^t-K,q]
\end{equation}

In the absence of super-potentials, an Abelian CSM theory has a flavor symmetry whose rank equal the number of chiral multiplets: gauging an $U(1)$ symmetry creates a new dual $U(1)$ symmetry carried by monopole operators. Superpotential 
terms will break some of the flavour symmetries. 
We can include the flavour symmetries in the notation by enlarging the matrices $K$ and $q$ to include the couplings to appropriate background connections. 

In Appendix \ref{app:monopole}, we review in detail the condition for a dressed monopole operator to be chiral and thus suitable to enter a super-potential. Schematically, a BPS monopole can only be dressed by chiral fields which are uncharged under the corresponding magnetic flux $m$. If we dress by $n_i \geq 0$ powers of each chiral multiplet, gauge-invariance imposes 
\begin{equation}
    \sum_{i|m \cdot q_i>0} (m \cdot q_i) q_i - K m = \sum_{i|m \cdot q_i=0} n_i q_i
\end{equation}
Notice that this formula is compatible with parity acting as $m \to -m$. We can denote such a dressed monopole operator as $\left(\prod_i \phi_i^{n_i}\right) V_m$.

\subsection{Basic dualities}

There is a trick which is often employed to ``derive'' IR dualities for ACSM theories: identify a piece of the theory which has a known IR dual description and plug that description into the original theory to get a candidate dual description for it. The basic assumption here is that we can decompose the overall RG flow to the far IR SCFT into an RG flow for the sub-theory followed 
by an RG flow for the full system. The candidate duality can be tested on protected quantities computable via localization, such as the super-conformal index or the ellipsoid partition function, or even on the simplified theories which arise from an Holomorphic-Topological twist \cite{Aganagic:2017tvx,Costello:2020ndc}. 

Assuming the validity of this manipulation and combining it with integrating in or out extra fields, one can usually generate very large orbits of conjectural dual ACSM theories with the same protected quantities. Conversely, it is often the case that different-looking ACSM theories have the same protected quantities and may be IR dual, but it can be challenging to find a chain of simpler manipulations relating them. 

The 3d-3d correspondence \cite{Dimofte:2011ju} offers a way to label certain duality orbits geometrically. The resulting daCSM theories are expected to describe the far IR physics of a 6d $(2,0)$ SCFT compactified on a three-manifold with defects. The 6d SCFTs are labelled by an ADE Lie algebra and the $A_1$ case is particularly well-studied. 

The dACSM theories which emerge from the 3d-3d correspondence have an useful feature: every super-potential term can be related by a chain of simple dualities to an order operator, a monomial of elementary fields. This type of theories was dubbed ``class ${\cal R}$'' in \cite{Dimofte:2011py}. We suspect our construction will produce class ${\cal R}$ descriptions of the 3d ${\cal N}=4$ SCFTs we are interested in even outside the geometric framework. 

\subsection{Integrating out fields}
As one goes through duality chains for a given dACSM theory, it is often possible to simplify the overall gauge group. For example:
\begin{itemize}
    \item Consider a situation with two $U(1)$ gauge fields $a_1$, $a_2$ with a minimal mixed CS coupling described by a $2 \times 2$ block 
    \begin{equation}
        \pm \begin{pmatrix} 0 & 1 \cr 1 & 0 \end{pmatrix} \, ,
    \end{equation}
    and no charged matter fields. Such fields can be integrated out,
    with the mixed CS couplings of other gauge fields $A_1$ and $A_2$ to $a_1$ and $a_2$ being converted to a mixed coupling between $A_1$ and $A_2$ by completing squares. 
    \item Consider a situation with an $U(1)$ gauge field $a$ with a minimal CS coupling $\pm 1$ and no charged matter fields. Such a gauge field gives an almost trivial contribution to the theory. Again, it can be integrated away. Mixed couplings to other gauge fields can be rearranged by completing squares.
\end{itemize}
More generally, we can integrate out safely any gauge fields which are not coupled to matter and have a $K$-matrix with determinant $\pm 1$. 

Another important manipulation involves a pair of chiral multiplets $\phi_1$, $\phi_2$
with opposite gauge charges, bilinear superpotential coupling $\phi_1 \phi_2$ and appearing at most linearly in other superpotential terms. These can be safely integrated away. A single chiral multiplet $\phi$ with quadratic superpotential $\phi^2$ and appearing at most linearly in other superpotential terms also gives an almost trivial contribution to a theory and can be integrated away. 

\subsubsection{The basic triality}
An important non-trivial duality involves an $U(1)_{\frac12}$ Chern-Simons theory coupled to a chiral multiplet of charge $1$, e.g. $\cT_{1,1}[1,1]$. This theory has chiral monopole operators of positive charge, with no dressing. In the IR, it is expected to flow back to a theory of a free chiral multiplet.

Keeping track of flavour symmetries, we can denote the IR duality as 
\begin{equation}
    \cT_{1,1}\left[\begin{pmatrix} 1&1\cr 1 &0\end{pmatrix},\begin{pmatrix}1\cr0 \end{pmatrix}\right] \simeq \cT_{0,1}[0,1] \, .
\end{equation}

By a parity transformation, $U(1)_{-\frac12}$ Chern-Simons theory coupled to a chiral also flows to a free chiral, but with charge $-1$ and an extra unit of background CS coupling:
\begin{equation}
    \cT_{1,1}\left[\begin{pmatrix} 0&1\cr 1 &0\end{pmatrix},\begin{pmatrix}1\cr0 \end{pmatrix}\right] \simeq \cT_{0,1}[1,-1] \, .
\end{equation}

The two versions of this duality, applied to individual chiral multiplets of a bigger theory, are employed in class ${\cal R}$ to present super-potential couplings as order operators. 

There is a (mostly well-defined\footnote{The action actually involves super-conformal boundary conditions for 4d $U(1)$ ${\cal N}=2$ gauge theory. Most boundary conditions can be identified as Neumann b.c. enriched by a 3d SCFT.}) $SL(2,\bZ)$ action on 3d SCFTs with an $U(1)$ global symmetry, where $T$ shifts the background CS coupling and $S$ gauges the global symmetry. In particular, $(ST)^3=1$. This duality implies that the $ST$ transformation leaves 
$\cT_{0,1}[0,1]$ invariant. 

This observation inspires a first example of how dualities can be chained up. Suppose that we 
replace the chiral in the $U(1)_{\frac12}$ CS theory by a second copy of the $U(1)_{\frac12}$ CS theory. The original gauge field has $K=1$ and is not coupled to any matter field. We can integrate it away. The new gauge field then gets $K=0$ and we recover the $U(1)_{-\frac12}$ Chern-Simons theory dual to the chiral multiplet. We can keep track of the global charges by enlarging the K-matrix to include background couplings: 
\begin{equation}
    \cT_{1,1}\left[\begin{pmatrix} 1&1\cr 1 &0\end{pmatrix},\begin{pmatrix}1\cr0 \end{pmatrix}\right] \simeq \cT_{2,1}\left[\begin{pmatrix} 1 & 1 & 1\cr 1 &1 &0 \cr 1 &0&0\end{pmatrix},\begin{pmatrix}0\cr1 \cr 0 \end{pmatrix}\right]  = \cT_{1,1}\left[\begin{pmatrix} 0 & -1 \cr -1 &-1\end{pmatrix},\begin{pmatrix}1 \cr 0 \end{pmatrix}\right]
\end{equation}
We see that the sign of the topological charge is flipped as expected, and there is the expected shift in the background CS coupling. 

As a second sample manipulation, consider again the $U(1)_{-\frac32}$ CS theory coupled to a chiral multiplet of charge $1$, aka $\cT_{1,1}\left[\begin{pmatrix} -1 & 1 \cr 1 &0\end{pmatrix},\begin{pmatrix}1 \cr 0 \end{pmatrix}\right]$. Replace the chiral multiplet by a second $U(1)_{\frac12}$ gauge field coupled to a dual chiral multiplet and integrate away the original gauge field: 
\begin{equation}
    \cT_{1,1}\left[\begin{pmatrix} -1 & 1 \cr 1 &0\end{pmatrix},\begin{pmatrix}1 \cr 0 \end{pmatrix}\right] \simeq \cT_{2,1}\left[\begin{pmatrix} -1 & 1 & 1\cr 1 &1 &0 \cr 1 &0&0\end{pmatrix},\begin{pmatrix}0\cr1 \cr 0 \end{pmatrix}\right]  = \cT_{1,1}\left[\begin{pmatrix} 2 & 1 \cr 1 &1\end{pmatrix},\begin{pmatrix}1 \cr 0 \end{pmatrix}\right]
\end{equation}
We have thus shown that the $U(1)_{\pm \frac32}$
CS theories are dual to each other, i.e. they have an emergent parity symmetry! More precisely, 
\begin{equation}
    \bcT_{1,1}\left[\begin{pmatrix} -1 & 1 \cr 1 &0\end{pmatrix},\begin{pmatrix}1 \cr 0 \end{pmatrix}\right] \simeq \cT_{1,1}\left[\begin{pmatrix} -1 & -1 \cr -1 &- 1\end{pmatrix},\begin{pmatrix}1 \cr 0 \end{pmatrix}\right]
\end{equation}
i.e. parity also flips the flavour charges and adds one unit of background CS coupling. 

\subsubsection{The XYZ duality}
Another important duality is the XYZ duality, which is closely related to the basic mirror symmetry for 3d ${\cal N}=4$ theories: SQED with two flavour, i.e. 
\begin{equation}
    \cT_{1,2}\left[ \begin{pmatrix}1 & 1 &0\cr 1 & 0&0\cr 0&0&0 \end{pmatrix},\begin{pmatrix}1 &-1\cr 0&0\cr0&-1\end{pmatrix}\right]\, ,
\end{equation}
flows in the IR to a theory of three chiral multiplets $X$, $Y$ and $Z$ with a cubic super-potential coupling $XYZ$. The identification maps $Z$ to the meson and $X$ and $Y$ to bare monopoles of magnetic charges $\pm 1$.

For later comparison, it is instructive to manipulate SQED by replacing the charge $-1$ chiral multiplet with an $U(1)_{-\frac12}$ CS theory. The result is a $U(1)^2$ CS theory with a charge $1$ chiral multiplet for each $U(1)$ factor and a mixed CS coupling:
\begin{equation}
    K= \begin{pmatrix} 0 & 1 \cr 1 &0\end{pmatrix}
\end{equation}
We can label the theory as
\begin{equation}
    \cT_{2,2}\left[ \begin{pmatrix}0& 1 &1&0\cr 1 & 0&0&1\cr 1&0&0&0 \cr 0&1&0&0\end{pmatrix},\begin{pmatrix}1 &0\cr 0&1\cr0&0 \cr 0&0\end{pmatrix}\right]\, ,
\end{equation}
Here the bare $(1,1)$ monopole is half-BPS, as well as $(-1,0)$ dressed by the chiral multiplet for the second factor, or vice-versa. These are the $X$, $Y$ and $Z$ operators in the $XYZ$ theory.

Notice that the original SQED theory had a charge conjugation symmetry. The dual $U(1)^2$ theory has a different $\bZ_2$ symmetry permuting the two gauge fields. These two symmetries generate the $S_3$ permutation symmetry visible in the $XYZ$ description. 

\subsection{Half-BPS boundary conditions and dualities}
We begin with a quick review of boundary conditions for pure Abelian CS theories. With a standard YM term in the gauge action and ${\cal N}=2$ SUSY, we can define half-BPS boundary conditions which restrict the gauge group $G$ to a sub-group $H$. The term Neumann is used to describe a situation where the full gauge group survives at the boundary, Dirichlet when the gauge group is fully broken. 

Such boundary conditions have a potential boundary anomaly given by the CS level $K_H$ restricted to $H$. This anomaly needs to be cancelled by boundary degrees of freedom. The simplest option is to take some purely chiral 2d matter $T_\partial$. This option is only available (unitarily) if $K_H \leq 0$. 

As a pure CS theory is topological, the boundary degrees of freedom will form a (relative) 2d chiral theory. Roughly, the boundary operators consist of boundary monopole operators with magnetic charge orthogonal to $H$, dressed by 2d chiral matter fields to cancel the $H$ gauge charges. The monopole operators have a spin which can become unbounded from below from certain ranges of values for $K$, signifying SUSY breaking. 

The option of Dirichlet b.c. only work if $K$ is positive-definite. Then the 2d ``theory'' is a free boson lattice chiral algebra of level $K$. This is the Abelian version of the famous WZW-CS theory correspondence. 

For Neumann b.c., instead, the 2d chiral algebra is the {\it coset} of $T_\partial$ by the $G$ current algebra of level $-K$. 

This discussion assumed a specific choice of chirality for the boundary conditions. The opposite choice is obtained from $K \to -K$. 

In the following, we will usually restrict ourselves to 
Dirichlet or Neumann b.c. for individual $U(1)$ factors of the gauge group.

Adding chiral matter requires us to pick boundary conditions for the 3d chirals as well. Simple options, again, are Neumann and Dirichlet. The latter can be deformed to force non-zero vevs for the chiral multiplets at the boundary, but only if the vev is gauge-invariant, i.e. fixed by $H$. The choice will kill boundary monopole operators for which chirals with non-zero vev feel a positive magnetic flux. 

In our conventions, a chiral multiplet with Dirichlet b.c. does not contribute to the boundary gauge anomaly, while a chiral with Neumann contributes $-1$. 

Super-potential terms can destabilize a boundary condition or require a deformation thereof. Roughly, this happens if the superpotential restricted to the boundary is non-vanishing. The boundary conditions can also be deformed by boundary E- and F- fermionic superpotential terms. 

The basic IR bulk dualities we discussed above can be extended to dualities for boundary conditions as well. Remarkably, simple and well-defined boundary conditions have explicit conjectural duals. Accordingly, one may attempt to define collections of dualities for boundary conditions of generic daCSM theories by applying these elementary dualities to sub-theories. The presence of interacting boundary conditions among the duality images complicate the matter but is not an obstruction. 
Geometric approaches involving four-dimensional manifolds are also possible. \cite{Dimofte:2019buf}

In order to discuss examples, it is useful to introduce some localization tools.

\section{Half-indices and quantum dilogarithms}
The half-index is the Witten index of the space of BPS boundary local operators, weighed by a twisted spin fugacity $q$ associated to a specific linear combination of spin and $R$-charge and, possibly, fugacities for flavour symmetries. 

For a free 3d chiral multiplet with Neumann b.c. and $R$-charge $0$, the half index takes the form
\begin{equation}
    I\!\!I^{\mathrm{chi}}_N(x;q) = \frac{1}{\prod_{n=0}^\infty (1-x q^n)} \equiv \frac{1}{(x;q)_\infty}
\end{equation}
counting the polynomials in the chiral multiplet scalar and its derivatives. 

For Dirichlet b.c. and $R$-charge $0$, the half index takes the form
\begin{equation}
    I\!\!I^{\mathrm{chi}}_D(x;q) = \prod_{n=0}^\infty (1-x^{-1} q^{n+1}) \equiv (x^{-1} q;q)_\infty
\end{equation}
counting the polynomials in a fermionic field and its derivatives. 

If we deform the Dirichlet b.c. by a boundary vev for the scalar field, we break the global symmetry and the half-index reduces to 
\begin{equation}
    I\!\!I^{\mathrm{chi}}_{D_c}(q) = ( q;q)_\infty
\end{equation}

There is a remarkable Nahm sum expansion for the Neumann half-index:  
\begin{equation}
    I\!\!I^{\mathrm{chi}}_N(x;q) = \frac{1}{(x;q)_\infty} = \sum_{n=0}^\infty \frac{x^n}{(q)_n}
\end{equation}
This is a first hint of an important duality: the Nahm sum can be interpreted as the half-index for a simple boundary condition for the $U(1)_{-\frac12}$ CS theory dual theory: Dirichlet for the gauge field (contributing the denominator below) and $D_c$ for the dual chiral multiplet (contributing the numerator below). The sum is over (negative) monopole charges. 
\begin{equation}
    I\!\!I_{D;D_c}^{\cT_{1,1}[0,1]}(x;q) =  \sum_{n=0}^\infty x^{-n} \frac{(q^{n+1};q)_\infty}{( q;q)_\infty}= I\!\!I^{\mathrm{chi}}_N(x^{-1};q) 
\end{equation}

There is also a Nahm sum expansion for the Dirichlet half-index: 
\begin{equation}
    I\!\!I^{\mathrm{chi}}_D(x;q) = (q x^{-1};q)_\infty = \sum_{n=0}^\infty (-1)^n x^{-n} \frac{q^{\frac{n^2+n}{2}}}{(q)_n}
\end{equation}
This Nahm sum can also be interpreted as the half-index of a $(D,D_c)$ boundary condition, now for the dual $U(1)_{\frac12}$ CS theory:
\begin{equation}
    I\!\!I_{D;D_c}^{\cT_{1,1}[1,1]}(x;q) =  \sum_{n=0}^\infty x^{-n} q^{\frac{n^2}{2}} \frac{(q^{n+1};q)_\infty}{( q;q)_\infty} = I\!\!I^{\mathrm{chi}}_D(- q^{\frac12}x;q) 
\end{equation}
The extra linear terms $(-q^{\frac12})^n$ arise from a re-definition of fermion number and R-symmetry. The $q^{\frac{n^2}{2}}$ factor arises directly from $K=1$.

These formulae are just the tip of an iceberg. For example, we can consider some variant of the second construction. \footnote{Other Dirichlet b.c. for $U(1)_{-\frac12}$ CS theory appear to break SUSY}
We can replace $D_c$ with $D$ for the chiral multiplet:
\begin{equation}
    I\!\!I_{D;D}^{\cT_{1,1}[1,1]}(x,y;q) =  \sum_{n=-\infty}^\infty x^{n} q^{\frac{n^2}{2}} \frac{(y^{-1} q^{1-n};q)_\infty}{( q;q)_\infty}
\end{equation}
and use a dualization on the chiral multiplet:
\begin{equation}
    I\!\!I_{D;D}^{\cT_{1,1}[1,1]}(x,y;q) =  \sum_{n=-\infty}^\infty\frac{x^{n} q^{\frac{n^2}{2}}}{( q;q)_\infty} \sum_{m=0}^\infty (-1)^m y^{-m} q^{- n m} \frac{q^{\frac{m^2+m}{2}}}{(q)_m}
\end{equation}
The first sum resums to a theta function $\theta(x q^{-m};q)$. This is analogous to integrating away the corresponding $U(1)_1$ gauge field, which leaves behind a boundary free fermion:
\begin{equation}
    I\!\!I_{D;D}^{\cT_{1,1}[1,1]}(x,y;q) = \sum_{m=0}^\infty \theta(x q^{-m};q)  (-1)^m y^{-m}\frac{q^{\frac{m^2+m}{2}}}{(q)_m}
\end{equation}
and the theta function periodicity finally gives 
\begin{equation}
    I\!\!I_{D;D}^{\cT_{1,1}[1,1]}(x,y;q) = \theta(x;q) \sum_{m=0}^\infty \frac{x^m (-q^{\frac12})^m y^{-m}}{(q)_m} = \theta(x;q) I\!\!I^{\mathrm{chi}}_N(-q^{\frac12}x y^{-1} ;\fq)
\end{equation}
This mathematical manipulation can be lifted to a 
``dualize a sub-theory proof'' that the $(D;D)$ boundary condition flows to a Neumann b.c. for the dual chiral dressed by a spectator 2d chiral fermion, represented by the $\theta$ function and arising from
integrating away a bulk gauge field with CS level $1$. \cite{Dimofte:2017tpi}

As a cautionary tale, we can look also at the $(D,N)$ case. It has a simple relation to the $(D,D)$ case, as one can convert a Dirichlet b.c. for the 3d chiral into a Neumann one by adding a boundary 2d chiral multiplet: 
\begin{equation}
    \frac{I\!\!I_{D;D}^{\cT_{1,1}[1,1]}(x,y;q)}{\theta(-y q^{-\frac12};q)} =  \sum_{n=-\infty}^\infty \frac{(-q^{\frac12})^n x^{n} y^{-n}}{( q;q)_\infty (y q^{n};q)_\infty}
\end{equation}
Redefining $x$ for convenience, we write 
\begin{equation}
    I\!\!I_{D;N}^{\cT_{1,1}[1,1]}(x,y;q) =  \sum_{n=-\infty}^\infty \frac{x^{n}}{( q;q)_\infty (y q^{n};q)_\infty}
\end{equation}
We can attempt a dualization:
\begin{equation}
    I\!\!I_{D;N}^{\cT_{1,1}[1,1]}(x,y;q) =  \sum_{n=-\infty}^\infty \frac{x^{n}}{( q;q)_\infty}\sum_{m=0}^\infty \frac{y^m q^{n m}}{(q)_m}
\end{equation}
but reversing the summation order is non-sensical. 

It is also instructive to look at Neumann b.c. for the gauge theories, starting from a further manipulation of the Nahm sums. For example, 
\begin{equation}
    I\!\!I^{\mathrm{chi}}_D(x;q) = \sum_{n=0}^\infty (-1)^n x^{-n} \frac{q^{\frac{n^2+n}{2}}}{(q)_n} = \sum_{n=-\infty}^\infty (-1)^n x^{-n} q^{\frac{n^2+n}{2}} \oint \frac{dz}{2 \pi i z} \frac{z^{-n}}{(z;q)_\infty}
\end{equation}
Doing the sum first, we find 
\begin{equation}
    I\!\!I^{\mathrm{chi}}_D(x;q) = (q;q)_\infty \oint \frac{dz}{2 \pi i z} \frac{\theta(-q^{-\frac12} x z;q)}{(z;q)_\infty}
\end{equation}
to be recognized as the half-index of a Neumann b.c. for the gauge field: the 
contour integral projects on $U(1)$-invariant operators, the prefactor counts gauginoes, the integrand includes the half-index for the Neumann b.c. 
as well as a 2d chiral fermion of gauge charge $1$. We should take $K=0$, so that the $1$ anomaly of the 2d fermion cancels the $-1$ boundary anomaly of the 
3d chiral:
\begin{equation}
    I\!\!I_{N;N}^{\cT_{1,1}[0,1]}(x;q) = (q;q)_\infty \oint \frac{dz}{2 \pi i z} \frac{\theta(x z;q)}{(z;q)_\infty}
\end{equation}
The 2d fermion anomaly also cancels the boundary anomaly in the coupling between the gauge and topological $U(1)$ symmetries, so that the b.c. preserves the topological $U(1)$. A specialization of the above formula relates $D_c$ for the 3d chiral multiplet and $(N,D)$ for the $U(1)_{-\frac12}$ theory. One can also argue that $(N,N)$ for the $U(1)_{\frac12}$ theory also flows to $D_c$ in the IR:
\begin{equation}
    I\!\!I_{N;N}^{\cT_{1,1}[1,1]}(x;q) = (q;q)_\infty \oint \frac{dz}{2 \pi i z} \frac{1}{(z;q)_\infty} = (q;q)_\infty
\end{equation}

\subsection{The pentagon identity as an half-index}
In the following we will often use variant of the free chiral half-indices where the scalar has been assigned R-charge $1$ and the indices are graded by $(-1)^R$ instead of $(-1)^F$, which is only possible for integrally-quantized $R$. It is convenient to introduce $\fq = -q^{\frac12}$ as well. 

We identify the Neumann half-index with the ``bosonic'' quantum dilogarithm
\begin{equation}
    E_q(x) \equiv \frac{1}{(-\fq x;q)_\infty}= \sum_{n=0}^\infty (-1)^n x^n\frac{\fq^n}{(q)_n}
\end{equation}
and the Dirichlet one with the ``fermionic'' quantum dilogarithm 
\begin{equation}
    D_q(x) \equiv (-\fq x^{-1};q)_\infty= \sum_{n=0}^\infty x^{-n} \frac{\fq^{n^2}}{(q)_n}
\end{equation}

The dilogarithms satisfy a famous pentagon identity which employs the quantum  torus algebra 
\begin{equation}
    x_{a,b} x_{c,d} = \fq^{ad-bc} x_{a+c,b+d}
\end{equation}
The fermionic identity reads
\begin{equation}
    D_q(x_{0,1})D_q(x_{1,0}) = D_q(x_{1,0}) D_q(x_{1,1}) D_q(x_{0,1}) 
\end{equation}
Inverting the expression and applying $x_{a,b} \to x_{-a,-b}$ gives the bosonic identity
\begin{equation}
    E_q(x_{1,0}) E_q(x_{0,1}) =  E_q(x_{0,1}) E_q(x_{1,1}) E_q(x_{1,0}) 
\end{equation}

In order to give the formula a 3d meaning, we can expand out and pick the coefficient of $x_{a,b}$. The bosonic formula gives:
\begin{equation}
    \frac{\fq^{ab}}{(q)_a (q)_b} = \sum_{c=0}^{\mathrm{min(a,b)}} (-1)^c \frac{\fq^{c^2-c-a b}}{(q)_{a-c} (q)_{b-c}(q)_c}
\end{equation}
which can be recombined into 
\begin{equation}
    \sum_{a=0}^\infty \sum_{b=0}^\infty x^a y^b \frac{q^{ab}}{(q)_a (q)_b} = \frac{(x y;q)_\infty}{(x;q)_\infty (y;q)_\infty}
\end{equation}
The right hand side has a natural interpretation as an half-index for the XYZ theory, with Neumann b.c for $X$ and $Y$ and Dirichlet for $Z$ (and R-charges $0$,$0$ and $2$ respectively, compatible with the $XYZ$ super-potential).

The left hand side is the half-index for a $(D,D;D_c,D_c)$ boundary condition of the dual $U(1)^2$ CS theory we introduced earlier on: the $\fq^{ab}$ comes from the mixed CS coupling and $m=(a,b)$ are the monopole charges. Either of the two sums can be done explicitly: 
\begin{equation}
    \sum_{a=0}^\infty \sum_{b=0}^\infty x^a y^b \frac{q^{ab}}{(q)_a (q)_b} =
     \sum_{a=0}^\infty x^a  \frac{1}{(q)_a (y q^a;q)_\infty} \, ,
\end{equation}
which is the half-index of a $(D;D_c,N)$ boundary condition for SQED. 

The fermionic formula gives 
\begin{equation}
    \frac{\fq^{a^2-ab+b^2}}{(q)_a (q)_b} = \sum_{c=0}^{\mathrm{min(a,b)}} \frac{\fq^{(a-c)^2 + (b-c)^2 +a b}}{(q)_{a-c} (q)_{b-c}(q)_c}
\end{equation}
which can be recombined into 
\begin{equation}
    \sum_{a=0}^\infty \sum_{b=0}^\infty (-1)^{a+b} x^{-a} y^{-b} \frac{\fq^{(a-b)^2+a+b}}{(q)_a (q)_b} = \frac{(q x^{-1};q)_\infty (q y^{-1};q)_\infty}{(q x^{-1} y^{-1};q)_\infty}
\end{equation}
The right hand side has a natural interpretation as an half-index for the XYZ theory, with Dirichlet b.c for $X$ and $Y$ and Neumann for $Z$.

The left hand side is the half-index for a $(D,D;D_c,D_c)$ boundary condition of the parity-reversed of the dual $U(1)^2$ CS theory we introduced earlier on. Either of the two sums can be done explicitly: 
\begin{equation}
    \sum_{a=0}^\infty \sum_{b=0}^\infty (-1)^{a+b} x^{-a} y^{-b} \frac{\fq^{a^2-2 a b + b^2+a+b}}{(q)_a (q)_b} =  \sum_{a=0}^\infty (-1)^a x^{-a} (q^{1-a} y^{-1};q)_\infty  \frac{\fq^{a^2+a}}{(q)_a} \, ,
\end{equation}
which is the half-index of a $(D;D_c,D)$ boundary condition for SQED. 

\subsection{Boundary duality webs.}
As an example, we can discuss boundary conditions for $\cT_{\mathrm{min}}$. Two natural choices are:
\begin{itemize}
    \item The $(N;D)$ boundary condition requires an extra 2d chiral fermion of charge $1$. This allows it to preserve the topological $U(1)_A$ symmetry. 
    \item The $(N;N)$ boundary condition can be arranged with two extra 2d chiral fermions, leading to a boundary condition which preserves $U(1)_A$ but also has an extra $\mathfrak{su}(2)_1$ boundary current algebra.
\end{itemize}
Dirichlet boundary conditions do not appear to viable. 

Next, we can replace the 3d chiral multiplet by an $U(1)_\frac12$ theory, with $(D;D_c)$ or $(D;D)$ boundary conditions. The latter option requires one to pair the 3d chiral with one of the two boundary fermions. 

We are then in position to integrate away the original $U(1)$ gauge field with Neumann b.c. coupled to a single 2d chiral fermion. This is a trivial theory with a trivial boundary. As a result, we arrive to a $\bcT_{\mathrm{min}}$ theory with $(D;D_c)$ or $(D;D)$ boundary conditions. Assuming the duality, we learn that the $(D;D)$ boundary condition must acquire in the IR an extra $\mathfrak{su}(2)_1$ boundary current algebra.

These manipulations are easily arranged at the level of the half-indices. The $(N;D)$ half-index is 
\begin{equation}
    I\!\!I^{\cT_{\mathrm{min}}}_{N;D}(x) = (q;q)_\infty \oint \frac{dz}{2 \pi i z} \theta(x z;q)(q z^{-1};q)_\infty = \sum_{n=0}^\infty \fq^n x^{n} \frac{q^{n^2}}{(q)_n} 
\end{equation}
with the sum being associated to the $(D;D_c)$ b.c.

The $(N;N)$ half-index is 
\begin{equation}
    I\!\!I^{\cT_{\mathrm{min}}}_{N;N}(x_1,x_2) = (q;q)_\infty \oint \frac{dz}{2 \pi i z} \frac{\theta(x_1 z;q)\theta(x_2 z;q)}{(z;q)_\infty} = \sum_{n=-\infty}^\infty x_2^{-n} x_1^n q^{n^2} \frac{(x_1^{-1} q^{\frac12-n};q)_\infty}{(q;q)_\infty}
\end{equation}
with the sum being associated to the $(D;D)$ b.c. We used
\begin{equation}
    \frac{\theta(x y;q)}{(x;q)_\infty} = \sum_{n=-\infty}^\infty x^n y^n q^{\frac{n^2}{2}} \frac{(y^{-1} q^{\frac12-n};q)_\infty}{(q;q)_\infty}
\end{equation}
derived above.

It is also useful to split out the decoupled $\mathrm{su}(2)_1$ characters by replacing the two chiral fermions with an $\mathrm{u}(1)_2$ lattice VOA, at the price of getting an interface from $\cT_{\mathrm{min}}$ to the $U(1)_2$ CS TFT instead of a boundary condition: 
\begin{equation}
    I\!\!I^{\cT_{\mathrm{min}}}_{N;N;\mathrm{u}(1)_2}(x) = \oint \frac{dz}{2 \pi i z} \frac{\sum_{n} q^{n^2} x^{2n} z^{2n}}{(z;q)_\infty} = \sum_{n=-\infty}^\infty x^{-2n} \frac{q^{n^2}}{(q)_{2n}}
\end{equation}

We have thus found two conjectural dual descriptions, one left and one right, for two $U(1)_A$-preserving boundary conditions of the minimal ${\cal N}=4$ SCFT. They are expected to be half-BPS in the IR and compatible with one of the two topological twists. 

Indeed, specializing $x = \fq$ in $I\!\!I^{\cT_{\mathrm{min}}}_{N;D}(x)$ gives the vacuum character of the ${\mathcal M}(2,5)$ minimal model chiral algebra. 
On the other hand, specializing  $x = \fq^{-1}$ in $I\!\!I^{\cT_{\mathrm{min}}}_{N;N;\mathrm{u}(1)_2}(x)$ gives the vacuum character of the ${\mathcal M}(3,5)$ minimal model chiral algebra. 
These are the expected specializations for the A- and B- twists. Specializing the half-index $I\!\!I^{\cT_{\mathrm{min}}}_{N;N}(x_1,x_2)$ gives the character of the $\mathrm{osp}(1|2)_1$ chiral algebra \cite{Ferrari:2023fez}.

\section{Schur index vs the 3d half-index}
The Schur index was originally introduced as a specialization of the super-conformal index of 4d ${\cal N}=2$ SCFTs. Contrarily to the full index, it can be defined in the absence of super-conformal symmetry. This motivates the existence of the ``IR formulae'' we use below, which compute the index in a Seiberg-Witten low-energy description of the theory. The Schur index coincides with the vacuum character of the protected chiral algebra subsector \cite{Cordova:2015nma,Cordova:2016uwk}. 

The IR formulae take as an input the BPS spectrum of the 4d theory. The spectrum depends on a choice of Coulomb vacuum $u$ of the 4d theory, but the output of the IR formula does not depend on this choice. The independence is a somewhat non-trivial consequence of certain wall-crossing formulae satisfied by the spectrum. 

In the cases at hand, the spectrum will consist of a collection of $2N$ BPS hypermultiplets with charges $\gamma_i$. The charges live in a lattice $\Gamma$ of gauge and flavour charges, with a sub-lattice $\Gamma_f$ of flavour charges. In the following we will need the integral Dirac inner products 
\begin{equation}
    \langle \gamma_i, \gamma_j \rangle = - \langle \gamma_j, \gamma_i \rangle \, ,
\end{equation}
between these charges. The inner product is linear and annihilates charges in $\Gamma_f$. We also need the gauge rank $\mathrm{rk}(\Gamma/\Gamma_f)$. 

The charges are given a specific cyclic order, with $\gamma_{i+N} = - \gamma_i$, determined by the phases of certain ``central charges'' $Z_{\gamma_i}$. The IR formula only uses the ordering information. 

The Schur index can be written formally as 
\begin{equation}
    I_{4d}(\mu;q) = (q)_\infty^{\mathrm{rk}(\Gamma/\Gamma_f)}\sum_{n_i}^{\sum_i n_i\gamma_i \in \Gamma_f} \frac{\mu^{\sum_i n_i \gamma_i} (-\fq)^{\sum_i n_i} \fq^{\sum_{i<j} \langle \gamma_i, \gamma_j\rangle n_i n_j} }{\prod_i(q)_{n_i}} \, ,
\end{equation}
where we picked any lift of the cyclic order to a linear order. This Nahm sum only makes sense for certain spectra, for which the quadratic exponent is positive in the $n_i \geq 0$ cone. 

If the sum makes sense, the Nahm sum will reproduce the character of the protected chiral algebra associated to the 4d SCFT. On the other hand, such a sum can be directly re-interpreted as the half-index of a $(D,D_c)$ boundary condition for some aCSM theory of rank $2N-\mathrm{rk}(\Gamma/\Gamma_f)$ and $K$ matrix determined by the quadratic exponent, with specialized flavour fugacities.

We would like to argue that, given appropriate super-potential terms compatible with the fugacity specialization, such a theory will flow to a 3d ${\cal N}=4$ SCFT and the $(D,D_c)$ boundary condition to  boundary condition compatible with the twist and leading to the same boundary chiral algebra.

In more general situations, the Schur index is computed by first assembling a ``spectrum generator'' $S$ from half of the BPS spectrum: 
\begin{equation}
    S_\gamma(q) = \sum_{n_i}^{\sum_i n_i\gamma_i =\gamma} \frac{(-\fq)^{\sum_i n_i} \fq^{\sum_{i<j} \langle \gamma_i, \gamma_j\rangle n_i n_j} }{\prod_i(q)_{n_i}} \, ,
\end{equation}
where $i=1, \cdots N$ and general properties of the central charges guarantee that the sum is finite for every $\gamma \in \Gamma$. For physical spectra, the smallest power of $q$ which survives in the sum is expected to grow with $\gamma$, potentially after heavy cancellations due to the signs in the sum. 

The index is then computed as
\begin{equation}\label{general schur}
    I_{4d}(\mu;q) = (q)_\infty^{\mathrm{rk}(\Gamma/\Gamma_f)}\sum_{\gamma,\gamma'}^{\gamma-\gamma'\in \Gamma_f} S_\gamma(q) S_{\gamma'}(q) \mu^{\gamma-\gamma'} \, .
\end{equation}
In such a situation, one may tentatively interpret $S_\gamma(q)$ as an half-index in an auxiliary 3d theory with a topological flavour symmetry $U(1)^{\mathrm{rk}(\Gamma)}$ 
and then associate $I_{4d}(\mu;q)$ to a 3d theory obtained by coupling two copies of the auxiliary theory to extra $U(1)$ gauge fields and, possibly, adding further super-potential deformations. 

If a 3d interpretation of this construction is available, it should be that the $(D,D_c)$ boundary condition cannot be directly defined in the naive daCSM theory, but it can be defined in a dual frame where we integrated in some extra 3d gauge fields. 

Another statement we would like to argue for is that the 3d theories and boundary conditions derived from different choices of 4d vacua $u$ for the same SCFT flow to the same SUSY-enhanced IR systems. 

In the following, we will learn how to compute other protected quantities
for these 3d theories and verify that they are all $u$-independent. 
In order to do so, we review a last piece of formalism. Define the quantum torus algebra
\begin{equation}
    X_{\gamma} X_{\gamma'} = \fq^{\langle \gamma, \gamma'\rangle} X_{\gamma + \gamma'} \, ,
\end{equation}
and a trace 
\begin{equation}
    \Tr X_\gamma = (q)_\infty^{\mathrm{rk}(\Gamma/\Gamma_f)} \delta_{\gamma,\Gamma_f} \mu^{\gamma} \, .
\end{equation}
Then, formally, 
\begin{equation}
    I_{4d}(\mu;q) = \Tr \prod_{i=1}^{2N} E_q(X_{\gamma_i}) \, .
\end{equation}
If necessary, this expression can be regularized by first computing 
\begin{equation}
    S = \prod_{i=1}^{N} E_q(X_{\gamma_i}) \, .
\end{equation}

\subsection{A quick example}
The so-called $(A_1,A_2)$ Argyres Douglas theory has a chamber of vacua such that the BPS spectrum consists of two particles only (and their anti-particles), with 
\begin{equation}
    \langle \gamma_1, \gamma_2\rangle = 1
\end{equation}
and no flavour charges.

Accordingly, 
\begin{equation}
    I_{\mathrm{4d}}(q) = (q)_\infty^2\sum_{n_i=0}^\infty \frac{q^{n_1 n_2+n_1 + n_2}}{(q)_{n_1}^2 (q)_{n_2}^2} 
\end{equation}
Experimentally, this coincides with the vacuum character of 
$M(2,5)$. A direct derivation of this fact is challenging, but possible by a sequence of identites we have already reviewed.\footnote{We thank Tomas Prochazka for a careful explanation.} 

As a first step, we rewrite 
\begin{equation}
    I_{\mathrm{4d}}(q) = (q)_\infty\sum_{n_i=0}^\infty \frac{q^{n_1 n_2+n_1 + n_2}}{(q)_{n_1}^2 (q)_{n_2}} (q^{n_2+1};q)_\infty
\end{equation}
and expand using a triality identity:
\begin{equation}
    I_{\mathrm{4d}}(q) = (q)_\infty\sum_{n_i=0}^\infty \frac{q^{n_1 n_2+n_3 n_2 +n_1 + n_2}}{(q)_{n_1}^2 (q)_{n_2}(q)_{n_3}} (-1)^{n_3} q^{\frac{n_3(n_3+1)}{2}}
\end{equation}
This makes the $n_2$ sum doable:
\begin{equation}
    I_{\mathrm{4d}}(q) = \sum_{n_i=0}^\infty (-1)^{n_3}q^{n_1} q^{\frac{n_3(n_3+1)}{2}}\frac{(q)_{n_1+n_3}}{(q)_{n_1}^2(q)_{n_3}} 
\end{equation}
Shifting a summation variable: 
\begin{equation}
    I_{\mathrm{4d}}(q) = \sum_{n=0}^\infty q^{n} (q)_{n} \sum_{c=0}^n (-1)^{c}  \frac{q^{\frac{c(c-1)}{2}}}{(q)_{n-c}^2(q)_{c}} 
\end{equation}
which by an XYZ identity
\begin{equation}
    I_{\mathrm{4d}}(q) = \sum_{n_=0}^\infty \frac{q^{n^2+n}}{(q)_{n}} 
\end{equation}
as desired. 

The original index can be identified with the $(D,D_c)$ half-index of a theory 
\begin{equation}\label{U(1) 2 description of Tmin}
    \cT_{2,4}\left[\begin{pmatrix}0 & 1 \cr 1 & 0\end{pmatrix},\begin{pmatrix}1 & 1 &0&0\cr 0&0 & 1&1\end{pmatrix}\right]
\end{equation}
The above manipulations have a natural interpretation as a chain of dualities. 
The first step dualizes one chiral multiplet to get an $U(1)^3$ theory. The corresponding $U(1)$ gauge field is now coupled to a single chiral multiplet and can be dualizable to a new chiral multiplet at the second step. The new chiral multiplet and the remaining original chiral multiplets are then subject to an XYZ duality (which would require the presence of a super-potential term), which erases a gauge field and a chiral multiplet. Two of the remaining chiral multiplets also appear to cancel against each other, which would require a second super-potential term. What is left is $\cT_{\mathrm{min}}$. 

We can follow this chain of manipulations back after turning on the $U(1)_A$ fugacity. Starting from  
\begin{equation}
    I_{\mathrm{4d}}(x;q) = \sum_{n_=0}^\infty \fq^n x^n \frac{q^{n^2}}{(q)_{n}} 
\end{equation}
we go to 
\begin{equation}
    I_{\mathrm{4d}}(x;q) = \sum_{n_=0}^\infty x^n \fq^{n} (q)_{n} \sum_{c=0}^n (-1)^{c}  \frac{q^{\frac{c(c-1)}{2}}}{(q)_{n-c}^2(q)_{c}} 
\end{equation}
and then 
\begin{equation}
    I_{\mathrm{4d}}(q) = (q)_\infty\sum_{n_i=0}^\infty \frac{q^{n_1 n_2+n_3 n_2 +n_1 + n_2}}{(q)_{n_1}^2 (q)_{n_2}(q)_{n_3}} x^{n_1+n_3} q^{\frac{n_3^2}{2}}
\end{equation}
which resums to 
\begin{equation}
    I_{\mathrm{4d}}(q) = (q)_\infty\sum_{n_i=0}^\infty \frac{q^{n_1 n_2 +n_1 + n_2}}{(q)_{n_1}^2 (q)_{n_2}} x^{n_1} (-\fq x q^{n_2};q)_\infty
\end{equation}
This suggests that $U(1)_A$ can be lifted to a combination of the topological symmetry 
at the second node (with minus sign) and the global symmetry of one of the chiral multiplets at the first node. This could also be re-defined to act only on the chiral multiplets at the first node, with opposite charge. 
Curiously, the same symmetry at the second node seems to flow to the same IR $U(1)_A$, at least based on the identity:
\begin{equation}
    \sum_{n_i=0}^\infty \frac{q^{n_1 n_2+n_1 + n_2}}{(q)_{n_1} (q)_{n_2}} x^{n_1} y^{n_2} (y q^{n_1+1};q)_\infty (x q^{n_2+1};q)_\infty = \sum_{n=0}^\infty (x y)^n \frac{q^{n^2+n}}{(q)_n}
\end{equation}
In fact, the $\CT_{2,4}$ description has four gauge invariant half-BPS monopole operators: 
\be
\phi_2 V_{(-1,0)}\ ,\quad \t\phi_2 V_{(-1,0)}\ ,\quad \phi_1 V_{(0,-1)}\ ,\quad \t\phi_1 V_{(0,-1)}\ ,
\ee
where $\phi_a$, $\t\phi_a$ are two chiral multiplets with charge 1 under the gauge group factor $U(1)_a$. We consider the following superpotential deformation
\be
W = \phi_2 V_{(-1,0)} + \t\phi_2 V_{(-1,0)}+ \phi_1 V_{(0,-1)}\ ,
\ee
which lifts the $U(1)^4$ flavour symmetry to a single $U(1)$ which rotates $\t\phi_2$. Performing F-maximization with respect to this $U(1)$, one can check that the superconformal index and the partition functions at the fixed point indeed agree with that of $\bcT_{\text{min}}$.

\subsection{The fermionic trace and parity-reversed boundary conditions}
It is natural to consider the quantity 
\begin{equation}
    D(\mu;q) = \Tr \prod_{i=2N}^{1} D_q(X_{\gamma_i})
\end{equation}
i.e. the trace of the inverse of the quantum torus algebra element which appears in the Schur index. This was studied before in \cite{Cecotti:2015lab}, together with traces of higher powers, and found to also be related to chiral algebra traces. A four-dimensional explanation of this fact is lacking. 

The fermionic dilogarithm products are much better behaved than the bosonic ones: no cancellations occur in wall-crossing formulae and thus if the expression converges for some vacuum it will converge in other vacua too.  

We have 
\begin{equation}
    D(\mu;q) = (q)_\infty^{\mathrm{rk}(\Gamma/\Gamma_f)}\sum_{n_i}^{\sum_i n_i\gamma_i \in \Gamma_f} \frac{\mu^{-\sum_i n_i \gamma_i} \fq^{\sum_i n^2_i} \fq^{\sum_{i>j} \langle \gamma_i, \gamma_j\rangle n_i n_j} }{\prod_i(q)_{n_i}}
\end{equation}
A simple interpretation is that this is the $(D,D_c)$ boundary condition for the parity-reversed 3d theory. 

Going back to the simple example, we can now execute some simple manipulations on the trace formula which were unsafe with bosonic dilogarithms. Starting from
\begin{equation}
    D(\mu;q) = \Tr D_q(X_{-\gamma_1}) D_q(X_{\gamma_2})D_q(X_{\gamma_1})D_q(X_{-\gamma_2})
\end{equation}
we apply the wall-crossing formula to 
\begin{equation}
    D(\mu;q) = \Tr D_q(X_{-\gamma_1})D_q(X_{\gamma_1})D_q(X_{\gamma_1+ \gamma_2}) D_q(X_{\gamma_2})D_q(X_{-\gamma_2})
\end{equation}
and recognize $D_q(x) D_q(x^{-1}) = \theta(x;q)$:
\begin{equation}
    D(\mu;q) = \Tr \theta(X_{\gamma_1};q)D_q(X_{\gamma_1+ \gamma_2}) \theta(X_{\gamma_2};q)
\end{equation}
leading immediately to 
\begin{equation}
    D(\mu;q) = \sum_{n=0}^\infty \frac{q^{n^2}}{(q)_n} 
\end{equation}
which indeed matches \begin{equation}
    D(\mu;q) =  (q)_\infty^2\sum_{n_i=0}^\infty \frac{q^{n_1^2- n_1 n_2+ n_2^2}}{(q)_{n_1}^2 (q)_{n_2}^2} 
\end{equation}
and is the specialization of the half-index at the ``wrong'' value of the $U(1)_A$ fugacity. 

Anther experimental formula (notice the permuted role of $x$ and $y$): 
\begin{equation}
    \sum_{n_i=0}^\infty \frac{q^{n_1^2- n_1 n_2+ n_2^2}}{(q)_{n_1} (q)_{n_2}} x^{n_1} y^{n_2} (x q^{n_1+1};q)_\infty (y q^{n_2+1};q)_\infty = \sum_{n=0}^\infty (x y)^n \frac{q^{n^2}}{(q)_n}
\end{equation}

\subsection{Super-conformal index formulae}
The trace $\Tr$ has positivity properties and can be used to promote the quantum torus algebra to an Hilbert space ${\mathcal H} \simeq \ell^2(\Gamma/\Gamma_f)$ equipped with a left- and right- action of the algebra itself as normal operators. This construction is based on the notion of Schur quantization \cite{Gaiotto:2024osr}, adapted to an IR description of the 4d theory \cite{upcoming}.

The operation of conjugation by $E_q(X_\gamma)$ defined an unitary operator $U_\gamma$ on ${\mathcal H}$:
\begin{equation}
    X_{\gamma'} \to E_q(X_\gamma)  X_{\gamma'} D_q(X_\gamma)
\end{equation}
i.e. a quantum cluster transformation:
\begin{equation}
    X_{\gamma'} \to X_{\gamma'} E_q(q^{\langle \gamma, \gamma'\rangle} X_\gamma)   D_q(X_\gamma)
\end{equation}
These operators satisfy pentagon identities, etc. 

We can then consider a trace $\Tr_{\mathcal H}$ of a product of $U_\gamma$ operators:
\begin{equation}
    I_{\mathrm{3d}} \equiv \Tr_{\mathcal H} \prod_{i=1}^{2N} U_{\gamma_i}
\end{equation}
Working through definitions in details, one discovers that this trace reproduces the 3d super-conformal index of the same aCSM theory we associated to the Schur index (with the same fugacity specialization). Thanks to the pentagon identities, this expression is wall-crossing invariant and independent of the choice of 4d vacuum!

Furthermore, the Hilbert space ${\mathcal H}$ has an alternative basis consisting of K-theory classes of line defects of the 4d theory \cite{Gaiotto:2024osr}. The combination of $\prod_{i=1}^{2N} U_{\gamma_i}$ acts on the line defects as a specific $U(1)_r$ R-symmetry rotation of the 4d theory. In the line defect basis, $I_{\mathrm{3d}}$ thus counts 4d line defects invariant under the R-symmetry rotation. We will discuss a physical explanation of this statement in later sections, but for now observe that this is enough to show $I=1$ in typical examples. This is expected for a theory with ${\cal N}=4$ IR enhancement and no Coulomb branch. 

We can give a sample calculation of how to employ wall-crossing identities to simplify an index calculation. Observe that the combination $U_\gamma U_{-\gamma}$ is a linear transformation on charges: 
\begin{equation}
    L_\gamma: X_{\gamma'} \to \theta(X_\gamma;q)  X_{\gamma'} \theta(X_\gamma;q)^{-1} =
     X_{\gamma'} \theta(q^{\langle \gamma, \gamma' \rangle} X_\gamma;q) \theta(X_\gamma;q)^{-1} = X_{\gamma'+\langle \gamma', \gamma \rangle \gamma}  
\end{equation}
Hence 
\begin{equation}
     I_{\mathrm{3d}} = \Tr_{\mathcal H} U_{-\gamma_1} U_{\gamma_2}U_{\gamma_1}U_{-\gamma_2}
\end{equation}
becomes
\begin{equation}
    I_{\mathrm{3d}} = \Tr_{\mathcal H} U_{-\gamma_1}U_{\gamma_1}U_{\gamma_1+ \gamma_2}U_{\gamma_2}U_{-\gamma_2}= \Tr_{\mathcal H} L_{\gamma_2} L_{\gamma_1}U_{\gamma_1+ \gamma_2}= \Tr_{\mathcal H} L_{\gamma_1} L_{\gamma_1 + \gamma_2} U_{\gamma_1+ \gamma_2}
\end{equation}
This is easily unpacked to the super-conformal index of $\cT_{\mathrm{min}}$ with specialized $U(1)_A$ fugacity.

\subsection{Teichm\"uller quantization and ellipsoid partition function}
Inspired by the super-conformal index formulae, we can replace the Hilbert space $\cH$ by $\widehat \cH = L^2(\bR^{\mathrm{rk}(\Gamma/\Gamma_f)/2})$, with an action of the Weyl algebra
\begin{equation}
    [x_{\gamma},x_{\gamma'}] = \frac{1}{2 \pi i}\langle \gamma,\gamma'\rangle\ .
\end{equation}

Accordingly, we can define 
\begin{equation}\label{general Sb}
    S_b(m) = \Tr_{\widehat \cH} \prod_{i=1}^{2N} \Phi_b (x_{\gamma_i})\ ,
\end{equation}
where $\Phi_b(x)$ is the Faddeev quantum dilogarithm. 
This is readily identified with the ellipsoid partition function of the usual 3d theory, now with specialized values for the mass parameters associated to flavour symmetries. 

The quantum dilogarithm satisfies the crucial pentagon identity
\begin{equation}
    \Phi_b(p)\Phi_b(x)=\Phi_b(x)\Phi_b(x+p)\Phi_b(p)
\end{equation}
if $[p,x] =(2 \pi i)^{-1}$. As a consequence, $S_b(m)$ is the same for 3d theories derived from the BPS spectrum in all 4d vacua.

The trace can be defined by writing the operator as an integral kernel acting on electric variables and then integrating over electric variables from $-\infty$ to $\infty$, fixing flavour variables to values $m$. More precisely, the algebra of $x_\gamma$
has a natural unitary representation on $L^2(\bR^{\mathrm{rk} (\Gamma/\Gamma_f)/2})$ via multiplication and shift operators and we take the trace over that Hilbert space. 

The quantum dilogarithm has good properties under Fourier transform, giving back the quantum dilogarithm with a shifted argument. (See appendix \ref{app: Faddeev} for various properties of the quantum dilogarithm.) This is related to the dualization operation on the chiral multiplet. For now, we can write
\begin{equation}
    \Phi_b(x_\gamma) = \int dp \,\hat \Phi_b(p) \, e^{2 \pi i p x_{\gamma}} 
\end{equation}
and then 
\begin{equation}\label{general ellipsoid}
    S_b(m)  = \int \prod_a \left[dp_a \hat \Phi_b(p_a)\right] e^{ \pi i \sum_{a<a'} \langle \gamma_a,\gamma_{a'}\rangle p_a p_{a'}} \delta_{\Gamma_f}\left(\sum_a \gamma_a p_a\right) e^{2 \pi i m \cdot \sum_a \gamma_a p_a}
\end{equation}
which is indeed the ellipsoid partition function for the 3d Chern-Simons theory we discussed, with some specific choices of masses/R-charge assignments. Again we emphasize that the order of the index $a$'s in the exponent are according to the phase of the central charges $Z(\gamma_a)$'s.

It should be possible to enrich the formula by turning on some other flavour fugacities which are present in the low-energy 3d effective description. We are particularly interested in the fugacity which corresponds to the extra R-charge generator for the extended SUSY algebra. 

The $(A_1,A_2)$ example can be studied as before. For convenience, let us denote $x_{\gamma_1} = p$ and $x_{\gamma_2} = x$. The ellipsoid partition function becomes 
\begin{equation}\label{ellipsoid minimal 1}
    S_b = \mathrm{Tr}\, \Phi_b(p) \Phi_b(x)\Phi_b(-p) \Phi_b(-x) 
\end{equation}

A single wall-crossing transformation gives 
\begin{equation}
    S_b = \mathrm{Tr}\, \Phi_b(x)\Phi_b(x+p)\Phi_b(p) \Phi_b(-p) \Phi_b(-x) 
\end{equation}
which simplifies (up to overall constants we disregard) to 
\begin{equation}
    S_b = \mathrm{Tr}\, e^{i \pi x^2}\Phi_b(x+p)e^{i \pi p^2}= \mathrm{Tr}\, e^{i \pi p^2}e^{i \pi x^2}\Phi_b(x+p)= \mathrm{Tr}\, e^{i \pi (x+p)^2}e^{i \pi p^2}\Phi_b(x+p)
\end{equation}
i.e. 
\begin{equation}\label{sigma twist Sb}
    S_b = \mathrm{Tr}\, e^{i \pi p^2}\Phi_b(x)e^{i \pi x^2} = \int \Phi_b(x)e^{i \pi x^2}dx
\end{equation}
which is the partition function of an $\bcT_{\text{min}}$. In particular, this expression computes the partition function in the topologically twisted limit and is independent of the parameter $b$.

It is possible to turn on the extra fugacity for the $U(1)_A := U(1)_C- U(1)_H$ symmetry in the IR by inserting an additional operator inside the trace:
\be\label{Sb refined AD2}
 S_b(y) = \mathrm{Tr}\, \Phi_b(p) \Phi_b(x)\Phi_b(-p) \Phi_b(-x) e^{2\pi i y x}
\ee
The same manipulation as above gives
\be
    S_b(y) = \int \Phi_b(x)e^{i \pi x^2 + 2\pi i x y}dx\ ,
\ee
which implies that the $U(1)_A$ symmetry can be identified with the topological symmetry of the $U(1)_{3/2}$ gauge theory, which is the only symmetry available assuming no superpotential terms.

Also, note that the formula \eqref{ellipsoid minimal 1} directly gives another dual description of $\bcT_{\text{min}}$:
\be
\CT_{4,4}\left[-\left(\begin{array}{cccc} 0 & 1 &0 & 1 \\ 1 & 0 & 1 & 0 \\0 & 1 &0 & 1 \\ 1 & 0 & 1 & 0 \end{array}\right), \left(\begin{array}{cccc} 1 & 0 & 0 & 0\\ 0 & 1 & 0 & 0 \\ 0 & 0 & 1 & 0 \\ 0 & 0 & 0 & 1 \end{array}\right)\right]\ .
\ee
Replacing each of the chiral multiplet by a $U(1)_{1/2}$ coupled to a chiral of charge 1, and integrating out the original gauge fields, we arrive at
\be
\CT_{2,4} \left[\left(\begin{array}{cc} 0 & 1 \\ 1 & 0 \end{array}\right), \left(\begin{array}{cccc} 1 & 1 & 0 & 0 \\ 0 & 0 & 1 & 1\end{array}\right)\right]\ ,
\ee
which is precisely \eqref{U(1) 2 description of Tmin}.

\subsection{Bethe equations}
The previous definitions have a neat classical limit. Define the classical cluster transformation 
\begin{equation}
    u_\gamma: X_{\gamma'} \to X_{\gamma'} (1+X_\gamma)^{\langle \gamma, \gamma'\rangle}
\end{equation}
This satisfies pentagon equations, etc. and is the classical ($b \to \infty$ or $q \to 1$) limit of the transformations induced by conjugation by a quantum dilogarithm.

We can define a set of equations in $(\bC^*)^{\mathrm{rk}(\Gamma/\Gamma_f)}$ 
by concatenating the transformations $u_{\gamma_i}$ from $i=1$ to $2N$ and looking for fixed points of the resulting transformation. We identify these with ``Bethe vacua'' for the associated 3d theory \cite{Nekrasov:2009uh}, i.e. vacua for the circle-compactified 3d theory, with specialized mass parameters associated to the flavour symmetries. 

The equations and their solutions can be manipulated as in the case of other protected quantities, demonstrating that the theories which arise from different chambers of the same 4d theory have the same Bethe vacua. Furthermore, we can demonstrate that the $(A_1,A_2)$ theory has the same Bethe vacua as $\cT_{\mathrm{min}}$, etc.

Finally, observe the $q \to 1$ limit of our statements about the superconformal index. The cluster transformations act naturally on 
certain Laurent polynomials (framed BPS generating functions) $F_L$ labelled by the half-BPS line defects of the 4d theory \cite{Gaiotto:2010be}. These are holomorphic functions on the ``K-theoretic Coulomb branch'' of the 4d theory. 

The overall sequence of transformations is a symmetry of the K-theoretic Coulomb branch, corresponding to an $U(1)_r$ rotation by $2 \pi$. 
As a consequence, Bethe vacua of the 3d theory are associated to fixed points of the $U(1)_r$ rotation in the K-theoretic Coulomb branch. 

Up to hyper-K\"ahler rotation, a direct relation to \cite{Fredrickson:2017yka,2017arXiv170906142F} appears very likely.
We leave a direct comparison to future work.

\section{A series of examples}

In this section, we discuss more examples of Argyres-Douglas theories of type $(A_1,G)$  for $G=A,D,E$. Let $r=\text{rk}(G)$, and $C(G)$ be the Cartan matrix of $G$. The ellipsoid partition function \eqref{general ellipsoid} and the Schur index \eqref{general schur} gives a general 3d gauge theory description in terms of $\CT_{2r,2r}$ theory coupled to a Lagrange multiplier multiplet which imposes the condition
\be
\sum_{i=1}^{2r}\gamma_i V_i = \gamma_fV_f\ ,
\ee
where $V_f$ is the background flavour multiplet. In the absence of the flavour symmetry ($G=A_{2N}$, $E_6$, and $E_8$), we also have a dual description 
\be\label{T r 2r description}
\CT_{r,2r}\left[ K_G, I_{r}\otimes (1,1)\right]\ ,
\ee
where the level matrix is given by
\be
(K_G)_{ab} = (2I-C(G))_{ab} = |\langle \gamma_a,\gamma_b\rangle|\ .
\ee
It is possible to check that the gauge theory \eqref{T r 2r description} contains the following gauge invariant half-BPS monopole operators:
\be
\left(\prod_{i} \phi_i^{|\langle \gamma_i,\gamma_j\rangle|} \right)V^-_j\ ,\qquad \left(\prod_{i} \t\phi_i^{|\langle \gamma_i,\gamma_j\rangle|} \right)V^-_j\ ,\qquad \forall i=1,\cdots r\ ,
\ee
where $V^-_{j}$ is the bare monopole operator with flux $m_{ja}=-\delta_{ja}$. The theory can be deformed by the superpotential containing all of the above monopole operators except for one. This lifts the $U(1)^{2r}$ flavour symmetry of the $\CT_{r,2r}$ theory to a single $U(1)$, which can be identified with $U(1)_A = U(1)_C-U(1)_H$ of the enhanced $\CN=4$ supersymmetry. 

The $(D,D_c)$ half-index of the $\CT_{r,2r}$ theory reads \footnote{The $q^{n\text{-linear}}$ factor in the numerator comes from a specific choice of $R$-charges of the chiral multiplets.}
\be
I_{G}(q) = (q)_\infty^{r} \sum_{n_1,\cdots,n_{r}=0}^\infty \frac{q^{\frac12\sum_{a,b} K(G)_{ab}n_a n_b + \sum_{a=1}^{r}n_a}}{\prod_{a=1}^{r} [(q)_{n_a}]^2 }\ ,
\ee
which reproduces the Schur index of the 4d theory.
One can check that the fermionic trace $D(q)$ is given by
\be
D_{G}(q) = (q)_\infty^{r} \sum_{n_1,\cdots,n_{r}=0}^\infty \frac{q^{\frac12\sum_{a,b} C(G)_{ab}n_a n_b }}{\prod_{a=1}^{r} [(q)_{n_a}]^2 }\ ,
\ee
which is indeed the $(D,D_c)$ half-index of the parity reversed theory
\be
\CT_{r,2r}\left[ C(G), I_{r}\otimes (1,1)\right] = \overline\CT_{r,2r}\left[ K_G, I_{r}\otimes (1,1)\right]\ .
\ee

Below we consider individual examples and discuss the relations to previous proposals in literature.

\subsection{$(A_1, A_{2N})$ theories}

Let us first consider the general $(A_1,A_{2N})$ AD theories, whose IR BPS spectrum in the canonical chamber is given by the charges $\gamma_i$'s, for $i=1,\cdots, 2N$ with
\be\label{canonical A2N}
\langle\gamma_i, \gamma_{i+1}\rangle = (-1)^{i+1}\ ,\quad i=1,\cdots, 2N-1\ .
\ee

\subsubsection{Ellipsoid partition functions}

Let us first consider the ellipsoid partition function, which reads
\be
S_b = \text{Tr}\prod_{a:\text{odd}}\Phi_b(x_a)\prod_{b:\text{even}}\Phi_b(x_b)  \prod_{a:\text{odd}}\Phi_b(-x_a)\prod_{b:\text{even}}\Phi_b(-x_b)\ .
\ee
This expression directly gives the description
\be
\CT_{4N,4N}\left[\t K_{4N}, I_{4N}\right]\ ,
\ee
where $I_{4N}$ is the $4N$-dimensional identity matrix and $\t K_{4N}$ is given by
\be
(\t K_{4N})_{ab} = \sum_{l=1}^N\sum_{k=0}^N (-1)^l(\delta_{a,2k+1}+ \delta_{a,2k+2N+1})(\delta_{b,2(k+l)}+\delta_{b,2(k+l+N)})\ .
\ee
Applying a particle-vortex duality for each of the chiral multiplet and integrating out the original gauge fields, we obtain another dual description
\be\label{A2N description 2}
\CT_{2N,4N} \left[K_{A_{2N}}, I_{2N}\otimes (1,1)\right]\ ,
\ee
where
\be
\left( K_{A_{2N}}\right)_{ab} = |\langle\gamma_a,\gamma_b\rangle|\ .
\ee
The gauge theory \eqref{A2N description 2} admits the following superpotential deformation:
\be\label{potential A2N}
W = \sum_{a=0}^{2N-1}\phi_a\phi_{a+2}V_{a+1} + \sum_{a=0}^{2N-2}\widetilde\phi_a\widetilde\phi_{a+2}V_{a+1}\ ,
\ee
where we set $\phi_0=\phi_{2N+1}=\t\phi_0=1$ and $V_{a}$ is the monopole operator with flux $(m_a)_i = -\delta_{ai}$.

Let $M_a$ and $\widetilde M_{a}$ for $a=1,\cdots 2N$ be the $U(1)$ global symmetry that rotates $\phi_a$ and $\widetilde\phi_a$ respectively. Then the superpotential \eqref{potential A2N} breaks the $U(1)_R\times U(1)^{4N}$ global symmetry to $U(1)_R\times U(1)_A$, with
\be
U(1)_A = \sum_{a=1}^N(-1)^{a+1} \t M_{2a+1}\ .
\ee
This leads to a conjectural $U(1)_A$-refined $S_b$ formula:
\be
S_b(y) = \text{Tr}\prod_{a:\text{odd}}\Phi_b(x_a)\prod_{b:\text{even}}\Phi_b(x_b)  \prod_{a:\text{odd}}\Phi_b(-x_a)\prod_{b:\text{even}}\Phi_b(-x_b)\prod_{b:\text{odd}} e^{2\pi i y(-1)^{(b-1)/2}x_b }\ .
\ee
Note that the last factor commutes with all $x_b$ except with $x_{2N}$.

\subsubsection{Bethe equations and other dual descriptions}

We claim that the theory \eqref{A2N description 2} together with the superpotential deformation \eqref{potential A2N} flows to the $\CN=4$ rank-zero theory $T_{N}$ introduced in \cite{Gang:2023rei}, which has a simpler description
\be\label{TNN description}
\CT_{N,N} \left[ K'_{N}, I_N\right]
\ee
with the superpotential 
\be
W = V_{m_1} + \cdots V_{m_{N-1}}\ ,
\ee
where $V_{m_i}$ is a bare monopole with the flux vector $m_i$ being the $i$-th row of the $C(A_N)$ matrix. The Chern-Simons coupling $K'_N$ in this description is
\be
{(K'_N)}_{ab} = 2(C(T_N)^{-1})_{ab} \ ,
\ee
where $C(T_N)^{-1}_{ab} = \text{min}(a,b)$ is the inverse of the Cartan matrix of the tadpole diagram, which is obtained by folding the $A_{2N}$ diagram in half.

As the first evidence for the duality, one can show that there exists a one-to-one correspondence between the two spaces of Bethe vacua. The Bethe equation for the $\CT_{2N,4N}$ description reads
\be\label{Bethe v}
\prod_{a=1}^{2N} u_a^{C(A_{2N})_{ab}} = (1-u_b)^2\ ,\qquad b=1,\cdots, 2N\ .
\ee
It turns out that all the solutions to the equation \eqref{Bethe v} satisfy the property
$u_{a} = u_{2N-a +1}\ ,~ a=1,\cdots N$.  
Imposing this relation, the system of equations reduces to that with $N$ variables, $u_1, \cdots u_N$. The reduced system of equations can be written as
\be
\prod_{a=1}^N u_a^{T_{ab}} = (1-u_b)^2\ ,
\ee
where $T_{ab} = C(T_N)_{ab}$.  Redefining $1-u_a = x_a$,
\be
1-x_a =\prod_{b} x_b^{2T^{-1}_{ab}}\ ,
\ee
which is precisely the Bethe equation for the $\CT_{N,N}$ description. One can also check for small values of $N$ that the twisted partition functions on Seifert manifolds and the superconformal indicies agree between the two descriptions.

It is possible to generate many other dual descriptions of the $T_N$ theory by performing wall-crossing manipulations. Let us illustrate an example for the $(A_1,A_{4})$ theory. We start from
\begin{equation}
    S_b = \mathrm{Tr}\, \Phi_b(x_1) \Phi_b(x_3)\Phi_b(x_2) \Phi_b(x_4) \Phi_b(-x_1) \Phi_b(-x_3)\Phi_b(-x_2) \Phi_b(-x_4) 
\end{equation}
with $\langle \g_1,\g_2\rangle = \langle \g_3,\g_2\rangle= \langle \g_3,\g_4\rangle=1$
and all others vanishing. 

A wall-crossing for $\g_3$ and $\g_2$ converts it into
\be
S_b =\text{Tr}\left[\Phi_b(x_1) \Phi_b(x_2)\Phi_b(x_2+x_3) \Phi_b(x_3)\Phi_b (x_4)\Phi_b(-x_1) \Phi_b(-x_3) \Phi_b(-x_2)\Phi_b (-x_4)\right]\ ,
\ee
which is the same as
\be
\text{Tr}\left[\Phi_b(-x_1)\Phi_b(-x_2)\Phi_b(x_1) \Phi_b(x_2)\Phi_b(x_2+x_3) \Phi_b(x_3)\Phi_b (x_4) \Phi_b(-x_3) \Phi_b (-x_4)\right]\ .
\ee
Another wall-crossing for $-x_2$ and $x_1$, and also for $x_4$ and $-x_3$ give
\be
S_b =\text{Tr}\left[ e^{i\pi x_1^2} \Phi_b(x_1-x_2) e^{i\pi x_2^2} \Phi_b(x_2 + x_3) e^{i\pi x_{x_3}^2} \Phi_b(-x_3+x_4) e^{i\pi x_4^2}\right]\ ,
\ee 
which is
\bea
S_b &=\text{Tr}\left[ e^{i\pi x_1^2} \Phi_b(x_1-x_2) \Phi_b(x_3) e^{i\pi x_2^2}e^{i\pi x_3^2} e^{i\pi x_4^2} \Phi_b( -x_3) \right] \\
&= \text{Tr}\left[ e^{i\pi x_1^2}  \Phi_b(x_1-x_2) \Phi_b(x_1-x_2 -x_3) e^{i\pi x_3^2}e^{i\pi x_2^2}e^{i\pi x_3^2} e^{i\pi x_4^2} \right] \\
&=  \text{Tr}\left[ e^{i\pi x_1^2}  \Phi_b(x_1-x_2) e^{i\pi x_3^2} e^{i\pi x_2^2}\Phi_b(x_1)e^{i\pi x_3^2} e^{i\pi x_4^2} \right] \\
&= \text{Tr}\left[  \Phi_b(-x_2)e^{i\pi x_1^2}  e^{i\pi x_3^2} e^{i\pi x_2^2}\Phi_b(x_1)e^{i\pi x_3^2} e^{i\pi x_4^2} \right] 
\eea 
This can be interpreted as $U(1)^2$ gauge theory with two chiral multiplets. It is convenient to identify
\be
(x_1,x_2,x_3,x_4) = (p_1,q_1,p_1+p_2,q_2)
\ee
with
\be
[p_i,q_j] = \frac{1}{2\pi i} \delta_{ij}\ .
\ee
Performing the Gaussian integrals, we arrive at
\be
S_b = \int dq_1 dp'_1 ~\Phi_b(q_1)\Phi_b(p'_1) e^{-\pi i {p'_1}^2 - 2 \pi i {q_1}^2 + 4\pi i q_1 p'_1 }\ ,
\ee
which is the ellipsoid partition function of
\be\label{A4 description 2}
\CT_{2,2}\left[\left(\begin{array}{cc} 0 & 2 \\ 2 & -1\end{array}\right),\left(\begin{array}{cc} 1 & 0 \\ 0 & 1\end{array}\right)\right]\ .
\ee 

One can check that this gauge theory allows a superpotential deformation $W = \phi_2^2 V_{(-1,0)}$, under which the theory is expected to flow to a $\CN=4$ rank-zero SCFT. 

Finally, we can turn on the extra fugacity for the $U(1)_A$ in the $S_b$ formula by considering
\begin{equation}\label{Sb refined AD4}
    S_b(y) = \mathrm{Tr}\, \Phi_b(x_1) \Phi_b(x_3)\Phi_b(x_2) \Phi_b(x_4) \Phi_b(-x_1) \Phi_b(-x_3)\Phi_b(-x_2) \Phi_b(-x_4) e^{2\pi iy (x_1-x_3)}\ .
\end{equation}
A similar wall-crossing manipulation gives
\be
S_b(y) =  \int dq_1 dp'_1 ~\Phi_b(q_1)\Phi_b(p'_1) e^{-\pi i {p'_1}^2 - 2 \pi i {q_1}^2 + 4\pi i q_1 p'_1 +2\pi i y q_1 -\pi i y^2 }\ .
\ee
This implies that the $U(1)_A$ can be identified with the topological symmetry for the second gauge group factor, which is indeed the only available flavor symmetry of the theory \eqref{A4 description 2} after the superpotential deformation. Calculation of the superconformal index and the partition functions at the fixed point strongly suggests that the theory is dual to the description \eqref{TNN description} for $N=2$,
\be
\CT_{2,2}\left[ \left(\begin{array}{cc} 2 & 2 \\ 2 & 4\end{array}\right),\left(\begin{array}{cc} 1 & 0 \\ 0 & 1\end{array}\right)\right]\ ,
\ee
with the superpotential deformation $W=V_{(2,-1)}$.

\subsubsection{Schur indices and the fermionic traces}

The $(D,D_c)$ half-index of the $\CT_{2N,4N}$ description \eqref{A2N description 2} directly reproduces the Schur index computed from the spectrum generator \cite{Cordova:2015nma}, 
\be
I_{A_{2N}}(q) = (q)_\infty^{2N} \sum_{n_1,\cdots,n_{2N}=0}^\infty \frac{q^{\frac12\sum_{a,b} (K_{A_{2N}})_{ab}n_a n_b + \sum_{a=1}^{2N}n_a}}{\prod_{a=1}^{2N} [(q)_{n_a}]^2 }\ ,
\ee
which coincides with the vacuum character of $M(2,2N+3)$. It is proven in \cite{Cecotti:2015lab} that this expression can also be written as
\bea
I_{A_{2N}}(q) &= \prod_{n\neq 0, \pm 1\text{mod} ~(2N+3)} (1-q^n)^{-1}\\
& = \sum_{n_1,\cdots, n_N=0}^\infty \frac{q^{\frac12\sum_{a,b} (K_{N}')_{ab}n_a n_b + \sum_{a=1}^{N} a n_a}}{\prod_{a=1}^{N} (q)_{n_a}}\ ,
\eea
which is indeed the $(D,D_c)$ half-index of the description \eqref{TNN description}.

The fermionic trace $D(q)$ of this theory is computed in \cite{Cecotti:2010fi}. In our notation, 
\be
D_{A_{2N}}(q) =(q)_\infty^{2N}\sum_{n_1,\cdots, n_{2N}=0}^\infty \frac{q^{\frac12\sum_{a,b}C(A_{2N})_{ab}n_a n_b}}{\prod_{a=1}^{2N}\left[(q)_{n_a}\right]^2}\ .
\ee
This can be interpreted as the $(D,D_c)$ half-index of
\be\label{2N 4N description}
\CT_{2N,4N} \left[C(A_{2N}), I_{2N}\otimes (1,1)\right]\ ,
\ee
which is indeed the parity conjugate description of \eqref{A2N description 2}. 

It is interesting to consider the linear chamber of the $(A_1,A_{2N})$ theory, which corresponds to the quiver with $2N$ nodes, with non-vanishing intersections 
\be
\langle \gamma_i,\gamma_{i+1}\rangle = 1\ , i=1,\cdots, 2N-1\ .
\ee
It is argued in \cite{Cecotti:2010fi} that the this is mutation equivalent to \eqref{canonical A2N}. Computation of the fermionic traces in this chamber gives rise to the following interesting $q$-series identity \cite{Cecotti:2010fi}
\be\label{fermionic trace A2N}
(q)_\infty^{2N}\sum_{n_1,\cdots, n_{2N}=0}^\infty \frac{q^{\frac12\sum_{a,b}C(A_{2N})_{ab}n_a n_b}}{\prod_{a=1}^{2N}\left[(q)_{n_a}\right]^2} = \sum_{n_1,\cdots,n_{2N-1}=0 }^\infty\frac{q^{\frac12\sum_{a,b}C(A_{2N-1})_{ab}n_a n_b}}{\prod_{a=1}^{2N-1}(q)_{n_a}}\ ,
\ee
which suggests another dual description of \eqref{2N 4N description}, which is
\be\label{partiy conjugate A2N}
\CT_{2N-1,2N-1}[C(A_{2N-1}), I_{2N-1}]\ .
\ee

This is precisely the description studied in \cite{Creutzig:2024ljv}. It is shown that the theory admits the superpotential deformation
\be
W = \sum_{a=1}^{2N-2}\phi_{a-1}\phi_{a+2}V_{m_a}\ , \quad (m_a)_b = \delta_{ab}+\delta_{(a+1)b}\ ,\quad \phi_0=\phi_{2N}=1\ ,
\ee
and the infrared theory is conjectured to be mirror dual to \eqref{TNN description}. In this case, the mirror dual coincides with the parity conjugate. Finally, the fermionic trace \eqref{fermionic trace A2N} can be identified with the specialized vacuum character of the chiral algebra $\mathrm{osp}(1|2N)_1$.

\subsection{$(A_1,A_3)$ theory}

Now we consider the simplest example which has a rank 1 Coulomb branch, with a single relevant coupling and an $SU(2)$ flavour symmetry. Turning on the relevant coupling, at $u=0$, we have a BPS particle of charge $\gamma_1$ and an $SU(2)$ doublet of BPS particles of charge $\gamma_2$ and $\gamma_3$, with $\langle \gamma_1, \gamma_{2} \rangle = \langle \gamma_3,\gamma_2 \rangle = 1$. The flavour symmetry corresponds to the charge vector
\be
\gamma_f = \gamma_1-\gamma_3\ .
\ee

\subsubsection{Ellipsoid partition function.}

The $S_b$ formula reads
\begin{equation}\label{A3 ellipsoid}
    S_b = \mathrm{Tr}\, \Phi_b(x_1) \Phi_b(x_3)\Phi_b(x_2)\Phi_b(-x_{1}) \Phi_b(-x_3)\Phi_b(-x_2) 
\end{equation}
We can apply one wall-crossing step
\begin{equation}
    S_b = \mathrm{Tr}\, \Phi_b(x_3) \Phi_b({x_2})\Phi_b(x_1+x_2) e^{i\pi x_1^2} \Phi_b(-x_3)\Phi_b(-x_2) \ ,
\end{equation}
which is 
\be
S_b = \Tr\, e^{i\pi x_2^2} \Phi_b(x_{2}+x_3) \Phi_b(x_3) e^{i\pi x_1^2} \Phi_b(x_2)\Phi_b(-x_3)\ .
\ee
Then
\be
S_b = \Tr\, e^{i\pi x_2^2} \Phi_b(-x_2) e^{i\pi (x_{2}+x_3)^2} e^{i\pi x_1^2} \Phi_b(x_2) \ .
\ee
Finally, 
\begin{equation}
    S_b = \mathrm{Tr}\,   e^{i \pi x_3^2}e^{i \pi x_2^2}  e^{i \pi x_1^2}e^{i \pi x_2^2} =\int dx dx' e^{-2i \pi (x-x')^2+ i \pi (x+m/2)^2+i \pi (x'-m/2)^2}
\end{equation}
i.e. 
\begin{equation}
    S_b = e^{\frac{2 i \pi}{3} m^2}
\end{equation}
This result is confusingly simple: this theory behaves in some ways as an $SU(2)$ Chern-Simons theory at level $\frac43$. 

One can also obtain another dual description directly from \eqref{A3 ellipsoid}. This formula gives a $U(1)^4$ gauge theory
\be
\CT_{4,6}\left[-\left(\begin{array}{cccc} 0 & 1 &0 & 1 \\ 1 & 0 & 1 & 0 \\0 & 1 &0 & 1 \\ 1 & 0 & 1 & 0 \end{array}\right), \left(\begin{array}{cccccc} 1 & 1 & 0 & 0 & 0 & 0\\ 0 & 0 & 1 & 0 & 0 & 0\\ 0 & 0 & 0& 1 & 1 & 0 \\ 0 & 0 & 0 & 0 & 0 & 1 \end{array}\right)\right]\ .
\ee
Applying the particle-vortex duality and integrating out the original gauge fields, we obtain a different dual $U(1)^4$ description:
\be\label{A3 description 2}
\CT_{4,6}\left[\left(\begin{array}{cccc} 0 & 1 & 0 & 0\\ 1 & 0 & 1 & 0 \\ 0 & 1 & 0 & 0\\0 & 0 & 0 & 0 \end{array}\right), \left(\begin{array}{cccccc} 1 & 0 & 0 & -1 & 0 & 0\\ 0 & 1 & 1 & 0 & 0 & 0\\ 0 & 0 & 0& 1 & 1 & 0 \\ 0 & 0 & 0 & 1 & 0 & 1 \end{array}\right)\right]\ .
\ee
The latter gauge theory admits a superpotential deformation 
\be
\phi_4\phi_5 V_{(0,-1,0,0)} + \phi_1\phi_3V_{(0,-1,0,0)} + (\phi_2+\tilde\phi_2) V_{(0,0,0,-1)}\ ,
\ee
which lifts the $U(1)^6$ flavour symmetry to $U(1)_f\times U(1)_A$, where we idenfity
\be
U(1)_f = U(1)_1^{\text{top}} - U(1)_3^{\text{top}}
\ee
with the flavor symmetry rotating $\gamma_1$ and $\gamma_3$ and 
\be
U(1)_A = U(1)_1^{\text{top}} + U(1)_3^{\text{top}}
\ee
with the part of the R-symmetry in the IR $\CN=4$ theory. 

One can explicitly calculate the the partition functions of the gauge theory on Seifert manifolds and reproduce the modula data of the chiral algebra $\mathrm{su}(2)_{-4/3}$. \footnote{In fact the Bethe equation has four solutions whereas the chiral algebra $\mathrm{su}(2)_{-4/3}$ has three admissible representations. The fourth solution evaluated on the handle-gluing operator diverges (See Appendix. \ref{app.seifert}). We obtain the correct modular $S$- and $T$-matrices once we remove the contribution from this solution. This is similar to what happens for the modular matrices constructed directly from the 4d $\CN=1$ description of the $(A_1,A_3)$ theory considered in \cite{Dedushenko:2018bpp}.}

\subsubsection{Schur index and the fermionic trace}

The $(D,D_c)$ half-index of \eqref{A3 description 2} directly reproduces the formula computed from the 4d spectrum generator \cite{Cordova:2015nma}:
\be
 I_{A_3}(q,x) = (q)_\infty^2\sum_{\substack{n_1,n_2,n_3\\ m_1,m_2,m_3}} \frac{q^{n_1n_2+n_2n_3}(-q^{1/2})^{n_1+n_2+n_3+m_1+m_2+m_3}}{(q)_{n_1}(q)_{n_2}(q)_{n_3}(q)_{m_1}(q)_{m_2}(q)_{m_3}} x^{n_1-m_1}\delta_{n_2,m_2}\delta_{n_1+n_3, m_1+m_3}\ .
\ee

Again, the fermionic trace takes the simplest expression in the linear chamber where 
\be
\langle \gamma_1,\gamma_2\rangle = \langle\gamma_2,\gamma_3\rangle=1\ . 
\ee
We have
\be
D_{A_3}(q,z) = \text{Tr}~ D_{q}(X_{\gamma_3})D_{q}(X_{\gamma_2})D_{q}(X_{\gamma_1})D_{q}(X_{-\gamma_3})D_{q}(X_{-\gamma_2})D_{q}(X_{-\gamma_1})\ ,
\ee
which can be manipulated into
\be
D_{A_3}(q,z) = \text{Tr}~\theta(X_{\gamma_3},q)D(X_{\gamma_2-\gamma_3};q)\theta(X_{\gamma_2},q)D(X_{\gamma_1-\gamma_2};q)\theta(X_{\gamma_1},q)\ .
\ee
Expanding each term and taking the trace,
\be
D_{A_3}(q,z) = \frac{1}{(q)_\infty}\sum_{k\in \mathbb{Z}} q^{k^2}z^k\sum_{l_1,l_2\geq 0} \frac{q^{\frac12 (2l_1^2+2l_2^2-2l_1l_2)+k(l_1+l_2)}}{(q)_{l_1}(q)_{l_2}}\ ,
\ee
which is the half-index of
\be
\CT_{3,2}\left[\left(\begin{array}{ccc} 2 & -1 & 1 \\ -1 & 2 & 1 \\ 1 & 1& 2\end{array}\right),\left(\begin{array}{ccc} 1 & 0 & 0 \\ 0 & 1 & 0 \\ 0 & 0& 0\end{array}\right)\right]\ .
\ee 
This theory admits one gauge invariant monopole operator $V_{(-1,-1,1)}$, which we can use to deform to flow to a rank-one $\CN=4$ SCFT.

\subsection{$(A_1,A_{2N+1})$ theories}

We start from the canonical chamber with $\gamma_{1,\cdots, 2N+1}$ satisfying
\be
\langle\gamma_i,\gamma_{i+1}\rangle = (-1)^{i+1}\ ,\quad i=1,\cdots, 2N\ .
\ee
The theory enjoys the $U(1)$ flavour symmetry for $N>1$, which corresponds to 
\be
\gamma_f = \sum_{i=0}^n (-1)^{i-1}\gamma_{2i+1}\ .
\ee
The ellipsoid formula gives the gauge theory description
\be\label{4N+2 description}
\CT_{4N+2, 4N+2}[K_{A_{2N+1}},I_{4N+2}]
\ee
together with a Lagrange multiplier gauge multiplet which imposes
\be
\sum_{i=1}^{4N+2}\gamma_i V_i = \left[\sum_{i=0}^n (-1)^{i-1}\gamma_{2i+1}\right]V_f\ ,
\ee
where $V_f$ is a background vector multiplet for the $U(1)$ flavour symmetry. The Chern-Simons level matrix is given by
\be
(K_{A_{2N+1}})_{ab} =  \frac12 |\langle\gamma_a,\gamma_b\rangle|\ ,
\ee
for $a,b=1,\cdots 4N+2$. The level is not properly quantized, but it reduces to an integer quantized matrix once we integrate out the Lagrange multiplier multiplet. As in the $(A_1,A_3)$ example, we anticipate that the theory admits a superpotential deformation which breaks the $U(1)^{4N+2}$ flavour symmetry to $U(1)_A\times U(1)_f$.

The $(D,D_c)$ half-index formula directly gives the 4d Schur index \cite{Cordova:2015nma}
\bea 
I_{A_{2N+1}}(q,z)=(q)^{2N}_\infty \sum_{\substack{n_1\cdots, n_{2N+1}\ ,\\ m_1,\cdots, m_{2N+1}=0}}^{\infty} &~\frac{(-q^{\frac12})^{\sum_{i=1}^{2N+1}(n_i+m_i)}q^{\sum_{j=1}^N m_{2j}(m_{2j-1}+m_{2j+1})}}{\prod_{i=1}^{2N+1} (q)_{n_i}(q)_{m_i}} \\
&~ \cdot [(-1)^{N+1}z]^{m_1-n_1} \prod_{i=1}^N \delta_{n_{2i},m_{2i}} \prod_{j=1}^N \delta_{(-1)^{j+1}n_1+n_{2j+1}, (-1)^{j+1}m_1+m_{2j+1}}\ ,
\eea
which is the vacuum character of the logarithmic $\CB_{N+2}$-algebra \cite{Creutzig:2017qyf}.

The fermionic trace $D(q,z)$ is computed in \cite{Cecotti:2010fi,Cecotti:2015lab}:
\be
D_{A_{2N+1}}(q,z) = \frac{1}{(q)_\infty}\sum_{k\in\mathbb{Z}} z^k q^{\frac12(N+1)k^2}\sum_{l_1,\cdots, l_{2N}=0}^\infty \frac{q^{\sum_{a,b} \frac12 C(A_{2N})_{ab}l_al_b -k\sum_{a:\text{even}} (-1)^{a/2} (l_{a-1}+l_{a}) }}{\prod_{a=1}^{2N}(q)_{l_{a}}}\ ,
\ee
which can be identified with the $(D,D_c)$ half index of the 3d theory
\be
\CT_{2N+1,2N} \left[\t K_{2N+1}, I_{2N}\right]\ ,
\ee
with the CS level given by
\be
(\t K_{2N+1})^{ab} m_a m_b = \sum_{a,b=1}^{2N}C(A_{2N})^{ab}m_am_b  -2\sum_{a:\text{even}}(-1)^{a/2}(m_{a-1}+m_a) m_{2N+1} + (N+1)m_{2N+1}^2\ .
\ee
This provides a simple parity conjugate description of \eqref{4N+2 description}.

\subsection{$(A_1,D_{2N+1})$ theory}

The last example is $(A_1, D_{2N+1})$ theory, whose BPS spectrum in the canonical chamber is given by the quiver with $\gamma_{1,\cdots, 2N+1}$ satisfying
\be
\langle \gamma_i,\gamma_{i+1}\rangle = (-1)^{i-1}\ ,\quad i=1,\cdots 2N-1\ ,\quad \langle \gamma_{2N-1},\gamma_{2N+1} \rangle = 1\ .
\ee
The theory enjoys the $SU(2)$ flavour symmetry which rotates the last two nodes. This corresponds to the charge vector
\be\label{flavor D 2N+1}
\gamma_f = \gamma_{2N+1}-\gamma_{2N}\ .
\ee

The ellipsoid formula \eqref{general ellipsoid} gives the description
\be
\CT_{4N+2,4N+2} [K_{D_{2N+1}}, I_{4N+2}]\ ,
\ee
with a Lagrange multipler multiplet that imposes the condition \eqref{flavor D 2N+1}. The CS level is again given by
\be
(K_{D_{2N+1}})_{ab} = \frac12 |\langle \gamma_a,\gamma_b\rangle|\ ,
\ee
for $a,b=1,\cdots, 4N+2$. As before, the level matrix reduces an integer matrix once we integrate out the Lagrange multiplier multiplet. We anticipate that the theory admits a superpotential deformation which breaks the $U(1)^{4N+2}$ flavor symmetry into $U(1)_A\times U(1)_f$, and the second factor enhances to $SU(2)_f$ in the IR.

After a small manipulation, the $(D,D_c)$ half-index of this description gives the formula
\bea
I_{D_{2N+1}}(q) = (q)^{2N}_\infty \sum_{\substack{m_1,\cdots, m_{2N+1}\ ,\\ n_{2N}, n_{2N+1}=0}}^\infty &~\frac{q^{\sum_{i=1}^{2N+1} m_i +  \frac12 \sum_{i,j=1}^{2N+1} b_{ij}^{D_{2N+1}}m_im_j}}{(q)_{k_{2N}}(q)_{n_{2N+1}}(q)_{m_{2N}}(q)_{m_{2N+1}}\prod_{i=1}^{2N-1}[(q)_{m_i}]^2} \\
&~\cdot z^{2(m_{2N+1}-n_{2N+1})} \delta_{n_{2N}+n_{2N+1}, m_{2N}+m_{2N+1}}\ ,
\eea
where $b_{ij}^{D_{2N+1}} = -C(D_{2N+1})_{ij} + 2 \delta_{ij}$, which coincides with the vacuum character of the chiral algebra $\mathrm{su}(2)_{-\frac{4N}{2N+1}}$ \cite{Cordova:2015nma}.




\section{R-flows and fake Janus domain walls}
A Coulomb vacuum $u$ of a 4d ${\cal N}=2$ SQFT is a vacuum whose effective low energy description is a pure Abelian gauge theory. We will denote as $\Gamma$ the lattice of gauge and flavour charges in a given vacuum. This locally constant as a function of $u$ and is equipped with a Dirac pairing $\langle \bullet, \bullet\rangle$. There is a sublattice $\Gamma_f$ if flavour charges annihilated by the pairing, as well as a quotient $\Gamma_g$ of gauge charges on which the pairing is non-degenerate.

The properties of a given Coulomb vacuum are controlled by the collection $Z_\gamma(u;\tau)$ of central charges. This is a complex linear function of the charge $\gamma \in \Gamma$ and depends holomorphically on the choice of vacuum $u$ and couplings $\tau$. In particular, the BPS bound implies that a particle of charge $\gamma$ has mass greater or equal to $|Z_\gamma(u)|$ in the vacuum $u$. The bound is saturated by BPS particles. By assumption, all $|Z_\gamma(u)|$ for populated charges must be non-zero in a Coulomb vacuum. 

As BPS particles can have electric and magnetic particles, they cannot be simultaneously included in an effective description of the system. BPS particles are classified by their spin. Particles with the lowest spin belong to hypermultiplets. We will focus on theories and vacua for which the BPS spectrum consists of a finite collection of hypermultiplets only.\footnote{There is a general expectation that BPS particles of higher spin, such as massive gauge bosons, will always be accompanied by infinitely many hypermultiplets.} 

A Janus configuration is defined by selecting a path $\tau(x^3)$ in the space of couplings of the 4d theory. It can be defined to preserve 3d ${\cal N}=2$ SUSY. A 3d SUSY vacuum for the system is given by a path $u(x^3)$ in the space of 4d vacua, such that the real part of $Z_\gamma(u(x^3);\tau(x^3))$ is constant. Locally, the map from $u$ to the real parts of central charges is one-to-one, so the path $u(x^3)$ is determined by the starting point and by $\tau(x^3)$.

As we move in the $x_3$ direction, 4d BPS particles feel an effective potential which is minimum at locations where $\mathrm{Im} \,Z_\gamma(u(x^3);\tau(x^3))=0$. We will say that the particles are ``trapped'' at these locations. Generically, BPS particles of non-collinear charges will be trapped at different locations. More precisely, the BPS particles have SUSY ground states at these locations, which behave as 3d chiral multiplets. Depending on the sign of $\partial_{x^3} \mathrm{Im}\,Z_\gamma(u(x^3);\tau(x^3))$, the trapped particles behave as 3d chiral multiplets of charge $\pm \gamma$. It is dressed by $\pm \frac12$ unit of background Chern-Simons coupling, depending on the sign of $\partial_{x^3} \mathrm{Arg}\,Z_\gamma(u(x^3);\tau(x^3))$.

At low energy and on a sufficiently generic trajectory in the space of 4d vacua, we can thus describe the Janus system as a 4d Abelian gauge theory coupled with 3d chiral multiplets at various $x_3$ locations. Superpotential couplings between the 3d chiral multiplets are expected to be created by instanton processes which allow particles to interact away from the wells. A 2d analogue of this phenomena was studied in \cite{Gaiotto:2015aoa}. We will treat the super-potential couplings as unknown quantities. 

We can implement electric-magnetic duality transformations somewhere between each location, so that the BPS particle is a standard electrically-charged particle in the local duality frame. Each duality transformation can be implemented by some Neumann interface with appropriate Chern-Simons couplings. 

As a result, the 4d fields will map to a large collection of 3d gauge fields with intricate Chern-Simons couplings and pairs of chiral fields for each BPS particle in the 4d Coulomb vacuum. One may further employ particle-vortex dualities to restructure the 3d presentation.
If we compute protected quantities of the resulting 3d system, we will find expressions such as products of quantum dilogarithms valued in the quantum torus algebra built from $\Gamma$ \cite{Cecotti:2011iy}. If the 4d theory has a class $S$ description \cite{Gaiotto:2009we}, the result is a class R theory \cite{Dimofte:2013lba} associated to some auxiliary three-manifold $M_3$.

If we combine the Janus system with a super-symmetric circle compactification, we get a configuration we dub a ``Janus loop''. Now we have a constraint on the initial and final $u$ along the closed path, which we expect to lead to a finite collection of
possible 3d vacua. Computing protected quantities will lead to appropriate traces of products of  quantum dilogarithms, akin to these we encountered in previous Sections. In a class $S$ situation, the manifold $M_3$ is a ``mapping cylinder''.

As we vary the choice of path $u(x_3)$, the collection of domain walls will change discretely, but the effective low energy description should change continuously. This  is expected to happen via 3d dualities. In particular, an elementary wall-crossing process should intertwine a 2-wall and a 3-wall configuration via an XYZ duality. 
If the path is swept across a generic singular locus where a 4d particle becomes massless, the Chern-Simons dressing of the corresponding 3d chiral will flip, but 
the evolution of the charge lattice $\Gamma$ will also change in a manner which compensates that. Quantum dilogarithm identities will insure that the protected quantities are ineed invariant. 

As a consequence, if we evolve the starting point of the path or $\tau(x_3)$, 
the effective 3d theory is expected to go through a chain of 3d dualities. 
In general, not all paths in the space of 4d vacua arise as Janus paths. 
On the other hand, the dualities associated to wall-crossing seem to imply that a 3d effective theory built from a ``fake'' Janus path $u(x_3)$ will be equivalent to 
that associated to an homotopy-equivalent true Janus path. This will certainly be true at the level of protected quantities. 

The reason we raise this point is that we would like to interpret the 3d theory associated to the Schur index as arising from a Janus loop in the space of 4d couplings and vacua of a 4d SCFT such that the phases of central charges rotate uniformly by an angle of $2 \pi$ around the loop. Such a loop is almost certainly not a true Janus loop. Nevertheless, it should produce a valid description of the 3d SCFT associated to the true Janus loop. A similar approach has been successful, for example, to describe Janus domain walls in class $S$ and build candidate 3d descriptions for the compactification of 6d SCFTs on three-manifolds \cite{Dimofte:2013lba}. 

\subsection{Super-symmetric $U(1)_r$-twisted circle compactifications}

There is a close relation between the Schur index and the Holomorphic-Topological twist of general 4d ${\cal N}=2$ SQFTs. The Schur index counts local operators in the HT twist of the theory. The relation to chiral algebras for SCFTs should arise from a further $\Omega$-deformation in the topological plane, localizing the operator algebra to a complex plane. 

We aim to make contact with the half-index of a 3d theory with at least ${\cal N}=2$ SUSY, possibly enhanced to ${\cal N}=4$. That quantity also admits a twisted description: it counts the local operators at an holomorphic boundary for the HT twist of the 3d theory, possibly deformed to a topological twist if the bulk has ${\cal N}=4$ SUSY. This is a natural context where to encounter a chiral algebra in 3d. 

Tentatively, we may replace the topological plane in 4d with a cigar geometry and compactify down to a 3d HT theory 
equipped with a holomorphic boundary condition. The deformation to a bulk topological twist arises naturally from the $\Omega$ deformation in 4d.

The main subtlety is that the 4d theory has a framing anomaly associated to $U(1)_r$ rotations in the original theory, so the circle compactification to 3d is twisted by a $2 \pi$ $U(1)_r$ rotation. This is trivial for a Lagrangian SCFT, but non-trivial for an Argyres-Douglas-type SCFT. For a non-conformal theory Lagrangian theory it corresponds to a $\theta$-angle rotation and thus to an extra 3d Chern-Simons coupling. This twist is apparently needed in order for a ``natural'' boundary condition to exist which lifts to a smooth 4d geometry. We will denote the twisted circle as $S^1_r$.

We can attempt to study $S^1_r$ compactification in a Coulomb vacuum of the 4d theory. This is the setting which justifies the IR formulae for the Schur index. The vacuum breaks conformal symmetry, but this is OK with the HT interpretation of the Schur index. 

Notice that the $S^3 \otimes_q S^1$ geometry associated to the Schur index can be re-interpreted as an HT geometry: a quotient $S_q$ of $\mathbb{C} \times \mathbb{R}^2$ by any scale transformation on $\mathbb{R}^2$ together by a rescaling of $\mathbb{C}$ by a factor of $q$. In this geometry and in a Coulomb vacuum, we expect BPS particles to be ``trapped'' at specific locations along the great circle corresponding to the origin of $\mathbb{C}$.

Following this intuition, the $S^1_r$ compactification involves BPS particles bound to specific locations of the
circle, determined by their central charges. Each trapped BPS particle gives rise to a 3d chiral multiplet, possibly equipped with a background Chern-Simons coupling.

Compared with the previous discussion, now the central charges conditions will be rotated by a phase $e^{2 \pi i x_3}$. We can re-define the couplings and Coulomb branch parameters to re-absorb this phase, so that the central charge conditions 
remain fixed but $u$ and couplings rotate by a $2 \pi$ $U(1)_r$ rotation around the circle. 

It would be interesting to describe the mapping cylinders $M_3$ for our examples, say in the language of \cite{Dimofte:2013lba}, and compare with other known geometric descriptions \cite{Gang:2018huc}. We leave that to future work.

\section{Line defects}
Half-BPS line defects in the 4d SCFT can be studies in a Coulomb vacuum, giving rise to the notion of framed BPS degeneracies and their generating functions $F_L$, which are Laurent polynomials in the quantum torus algebra associated to $\Gamma$ \cite{Gaiotto:2010be}. 

The Schur index can be generalized to count local operators at the end of a line defect. The resulting $I_L(q)$ can be computed as 
\begin{equation}
    I_{L}(\mu;q) = \Tr F_L\prod_{i=1}^{2N} E_q(X_{\gamma_i}) \, .
\end{equation}
Expanding out, if we denote the coefficients of $X_\gamma$ in $F_L$ as $F_{L,\gamma}$, we get 
\begin{equation}
    I_{L}(\mu;q) = (q)_\infty^{\mathrm{rk}(\Gamma/\Gamma_f)}\sum_\gamma F_{L,\gamma} \left[\sum_{n_i}^{\gamma + \sum_i n_i\gamma_i \in \Gamma_f} \frac{\mu^{\gamma + \sum_i n_i \gamma_i} (-\fq)^{\sum_i n_i} \fq^{\sum_{i<j} \langle \gamma_i, \gamma_j\rangle n_i n_j} }{\prod_i(q)_{n_i}} \right]\, ,
\end{equation}
It is not difficult to interpret this expression as an half-index. Intuitively, the 4d line defect maps to a direct sum of 3d line defects, 
with multiplicities $F_{L,\gamma}$, each of which is a Wilson line for the auxiliary gauge fields valued in $\Gamma$. 

There are analogous modifications available for the trace formulae for the superconformal index and ellipsoid partition function of the 3d theory. 

In the former case, the map
\begin{equation}
    U_{L,L'} = X_{\gamma} \to F_L X_\gamma F_{L'} \, ,
\end{equation}
can be inserted in the formula:
\begin{equation}
    I^{L,L'}_{\mathrm{3d}} \equiv \Tr_{\mathcal H} U_{L,L'}\prod_{i=1}^{2N} U_{\gamma_i} \, .
\end{equation}
We interpret this as a 3d index decorated by two line defects, counting local operators at a junction of the two. 

In the latter, we can consider insertions of $F_L(e^{2 \pi i b x_\gamma})$ or of $F_L(e^{2 \pi i b^{-1} x_\gamma})$:
\begin{equation}
    S^{L,L'}_b(m) = \Tr_{\widehat \cH} F_L(e^{2 \pi i b x_\gamma}) F_{L'}(e^{2 \pi i b^{-1} x_\gamma}) \prod_{i=1}^{2N} \Phi_b (x_{\gamma_i})\ .
\end{equation}
We interpret this as an ellipsoid partition function decorated by two line defects, wrapping distinct supersymmetric circles in the geometry. 
At $b=1$, such expression is expected to compute the $S_{L,L'}$ Hopf link correlation function of line defects in the topologically-twisted 3d theory. It would be interesting to develop this point further.

Finally, the expectation value of circle-wrapped line defects int he Bethe vacua is computed by evaluating the $q \to 1$ limit of $F_L$ onto the fixed points $X_\gamma$.

\acknowledgments
This research was supported in part by a grant from the Krembil Foundation. DG
is supported by the NSERC Discovery Grant program and by the Perimeter Institute
for Theoretical Physics. Research at Perimeter Institute is supported in part by the
Government of Canada through the Department of Innovation, Science and Economic
Development Canada and by the Province of Ontario through the Ministry of Colleges
and Universities. The work of HK is supported by the Ministry of Education of the Republic of Korea and the National Research
Foundation of Korea grant NRF-2023R1A2C1004965.
\appendix

\section{Monopoles and super-potentials for ${\cal N}=2$ Abelian Chern-Simons matter theories} \label{app:monopole}
We use the conventions where each chiral multiplet is accompanied by $-\frac12$ unit of background CS coupling. 
Denote as $K$ the matrix of CS couplings on top of that and as $q_i$ the charges of the $i$-th chiral multiplet. Initially, we turn off any super-potential.

An half-BPS monopole operator of magnetic charge $m$ can only be dressed by chiral multiplets with $m \cdot q =0$. 
Gauge-invariance requires 
\begin{equation}
    K m +\sum_{i|m \cdot q =0}  n_i q_i= \sum_{i|m \cdot q >0} (m \cdot q_i) q_i
\end{equation}
with $n_i \geq 0$, the power of the chirals dressing the bare monopole. Positive multiples of a solution are a solution: the gauge-invariant monopoles can be raised to any power and thus parameterize a $\bC$ branch of vacua. 

The R-charge of these monopole operators depends on the R-charges $r_i$ of the chiral multiplets: 
\begin{equation}
   R= \rho \cdot m + \sum_{i|m \cdot q =0}  n_i r_i + \sum_{i|m \cdot q >0} (m \cdot q_i) (1-r_i)
\end{equation}
We included a mixing $\rho$ with topological charges. When adding operators to the super-potential, they should have $R$-charge $2$. 

For example, in $U(1)_{-\frac12}$ CS theory with a chiral of charge $1$, the monopole charge must be negative and the basic monopole has $R$-charge $-\rho$. In $U(1)_{\frac12}$ CS theory with a chiral of charge $1$ the monopole charge must be positive and the basic monopole has $R$-charge $-\rho+1-r_i$.
For other choices of level, the bare monopole is not gauge-invariant. 

\subsection{Rank two examples}
Next, consider an $U(1)\times U(1)$ gauge theory, with each factor coupled to a separate chiral multiplet of charge $1$. 

First, consider a $K$ matrix 
\begin{equation}
    K = \begin{pmatrix}
        2 & 2 \cr 2 & 4
    \end{pmatrix}
\end{equation}
which is expected to lead to a SUSY enhancement with a monopole super-potential of charge $(2,-1)$. The $K$ matrix produces charges $(2,0)$ which are cancelled by the last term in the constraint, without any dressing. This appears to be the only possible half-BPS monopole. It preserves an $U(1)_A$ symmetry built from the topological charges with $(1,2)$ coefficients. 

In the main text we also encounter 
\begin{equation}
    K = \begin{pmatrix}
        0 & 2 \cr 2 & -1
    \end{pmatrix}
\end{equation}
This allows for a monopole of charge $(-1,0)$ dressed by two copies of the second chiral multiplet. 

Finally, we encounter 
\begin{equation}
    K = \begin{pmatrix}
        0 & 2 \cr 2 & 0
    \end{pmatrix}
\end{equation}
coupled a pairs of chirals for each factor. Here the bare $(1,1)$ monopole is half-BPS, as well as $(-1,0)$ dressed by two chiral multiplet for the second factor, or vice-versa. 
The latter operators transform as triplets of the $SU(2)$ global symmetry acting on each pair of chirals. We should thus be able to preserve a Cartan of both. 

\section{Faddeev quantum dilogarithm identities} \label{app: Faddeev}
We have the pentagon identity:
\begin{equation}\label{pentagon}
    \Phi_b(p) \Phi_b(x) =\Phi_b(x) \Phi_b(x+p) \Phi_b(p) 
\end{equation}
if $[p,x]=(2 \pi i)^{-1}$. We also have $\Phi_b(0) = e^{\frac{\pi i}{24}(b^2 + b^{-2})}$ and
\begin{equation}\label{two-term}
    \Phi_b(x)\Phi_b(-x) = \Phi_b(0)^2 e^{\pi i x^2}
\end{equation}
When $b>0$, $\Phi_b(-\infty)=1$.

If we write 
\begin{equation}
    \Phi_b(p) = \int dy \,\hat \Phi_b(y) \, e^{2 \pi i y p} 
\end{equation}
and bring the $p$ exponential to the right, we get 
\begin{equation}\label{pentagon}
    \hat \Phi_b(y_1)\Phi_b(x+y_1) =\Phi_b(x) \int dy_2 \,\hat \Phi_b(y_2) \, e^{\pi i y_2^2} e^{2 \pi i y_2 x} \,\hat \Phi_b(y_1-y_2) 
\end{equation}
or 
\begin{equation}\label{pentagon}
    \hat \Phi_b(y_1)\Phi_b(x+y_1) \Phi_b(-x) =\Phi_b(0)^2 \int dy_2 \,\hat \Phi_b(y_2) \, e^{\pi i (x+y_2)^2} \,\hat \Phi_b(y_1-y_2) 
\end{equation}

then the pentagon identity can be decomposed into individual exponentials of $p$ and $x$:
\begin{equation}\label{pentagon2}
    e^{\pi i y_1 y_2} \hat \Phi_b(y_1)\hat \Phi_b(y_2) = \int dy \,\hat \Phi_b(y_2-y)\hat \Phi_b(y) \hat \Phi_b(y_1-y)e^{-\pi i (y_2-y) y-\pi i (y_2-y) (y_1-y)-\pi i y(y_1-y)}
\end{equation}
e.g. 
\begin{equation}\label{pentagon2}
    e^{2\pi i y_1 y_2} \hat \Phi_b(y_1)\hat \Phi_b(y_2) = \int dy \,\hat \Phi_b(y_2-y)\hat \Phi_b(y) \hat \Phi_b(y_1-y)e^{\pi i y^2}
\end{equation}

\section{Partition functions on Seifert manifolds}\label{app.seifert}

Let us consider a class of Seifert manifold labeled by two integers, $M_{g,p}$, which is a degree $p$ circle bundle over a genus $g$ Riemann surface. The partition function can be written as \cite{Closset:2017zgf}
\be\label{seifert bethe}
Z_{g,p} = \sum_{\{u_\beta:P_a(u_\beta)=1\}} H^{g-1}(u_\beta) F^p(u_\beta)\ ,
\ee
where the sum is taken over the solutions to the equations
\bea
P_a(u) = \exp\left[2\pi i\frac{\partial W_{\text{eff}}}{ \partial u_a}\right] = 1\ ,\quad \forall a\ ,
\eea
where $W_{\text{eff}}$ is the twisted effective superpotential of 2d $\CN=(2,2)$ theory on the Riemann surface. We defined the fibering operator
\be
F(u) = \exp\left[2\pi i \left(W_{\text{eff}} - u_a \partial_{a} W_{\text{eff}}\right)\right]
\ee
and the handle-gluing operator
\be
H(u) = e^{2\pi i\Omega(u)} \det_{ab} \partial_a \partial_b W_{\text{eff}}(u)\ ,
\ee
where $\Omega$ is the effective dilaton, which determines the coupling to the twisted background on the Riemann surface.

In the topological $A/B$-twisting limit, the partition function can be constructed from the modular $S$ and $T$-matrices of the boundary chiral algebra:
\be
Z_{g,p} = \sum_{P_a(u_\beta)=1} S_{0\beta}^{2-2g} T_{\beta\beta}^p\ ,
\ee
where the vacuum equation $P_a(u)=1$ now corresponds to the fusion rules of line operators in the 3d topological field theory. This formula allows us to identify $\{|S_{0\alpha}|^2\}$ and  $\{T_{\alpha\alpha}\}$ with $\{H(u_\alpha)^{-1}\}$ and $\{F(u_\alpha)\}$ up to an overall phase. The formula \eqref{seifert bethe} has an integral expression
\be\label{integral seifert}
Z_{g,p} = \frac{1}{|W_G|}\sum_{{\mathfrak m}\in \Gamma^{(p)}_{G^\vee}}\int du~ \left(\det_{ab} \partial_a \partial_b W_{\text{eff}}(u)\right)^g~ I_{g,p} (u,{\mathfrak m})\ ,
\ee
where $I_{g,p}$ is the contribution from the classical and one-loop integral in the presence of the torsion flux ${\mathfrak m}$. For example, a chiral multiplet $\phi$ of charge 1 under $G=U(1)$ (with R-charge 1) contributes
\be
I^\phi_{g,p}(u,{\mathfrak m}) = (1-e^{2\pi i u})^{-{\mathfrak m}}\exp \left[\frac{p}{2\pi i} \text{Li}_2(e^{2\pi i u}) + p u \ln (1-e^{2\pi i u})\right] \ , 
\ee
where ${\mathfrak m}\in \mathbb{Z}_p$. ($g$ dependence does not appear since we set $R$-charge 1.) Similarly to the quantum dilogarithm in the ellipsoid partition function, $I^\phi_{g,p}(x_\gamma,{\mathfrak m})$ can be interpreted as the operator acting on the Hilbert space $L^2(\mathbb{R})\otimes \mathbb{Z}_p$, where the inner product is defined by the integration measure in \eqref{integral seifert}. Indeed, this function is a natural generalization of the quantum dilogarithm at $b=1$ \cite{Dimofte:2014ija,Alexandrov:2015xir}, i.e., if we restrict to $p=1$, $I^\phi(u)$ reduces to $\Phi^{-1}_{b=1}(-i u)$. This function satisfies a pentagon relation and a two-term identity similar to the quantum dilogarithm \cite{Alexandrov:2015xir}.

\bibliographystyle{nb}
\bibliography{jobname}

\end{document}